\documentclass{article}
\usepackage{amsmath,mathtools}

\newcommand{\diagentry}[1]{\mathmakebox[1.8em]{#1}}
\newcommand{\xddots}{%
  \raise 4pt \hbox {.}
  \mkern 6mu
  \raise 1pt \hbox {.}
  \mkern 6mu
  \raise -2pt \hbox {.}
}
\newcommand{\rvline}{\hspace*{-\arraycolsep}\vline\hspace*{-\arraycolsep}}

\usepackage{alltt}
\usepackage[vcentermath]{youngtab}
\usepackage[all]{xy}
\usepackage{latexsym,amssymb,amsmath,amsfonts, amsthm, xcolor}

\newcommand{\CC}{\mathbb{C}}

\def\vbar{\mathchoice{\vrule height6.3ptdepth-.5ptwidth.8pt\kern-.8pt}
  {\vrule height6.3ptdepth-.5ptwidth.8pt\kern-.8pt}
  {\vrule height4.1ptdepth-.35ptwidth.6pt\kern-.6pt}
  {\vrule height3.1ptdepth-.25ptwidth.5pt\kern-.5pt}}
\def\fudge{\mathchoice{}{}{\mkern.5mu}{\mkern.8mu}}
\def\bbc#1#2{{\rm \mkern#2mu\vbar\mkern-#2mu#1}}
\def\bbb#1{{\rm I\mkern-3.5mu #1}}
\def\bba#1#2{{\rm #1\mkern-#2mu\fudge #1}}
\def\bb#1{{\count4=`#1 \advance\count4by-64 \ifcase\count4\or\bba A{11.5}\or
  \bbb B\or\bbc C{5}\or\bbb D\or\bbb E\or\bbb F \or\bbc G{5}\or\bbb H\or
  \bbb I\or\bbc J{3}\or\bbb K\or\bbb L \or\bbb M\or\bbb N\or\bbc O{5} \or
  \bbb P\or\bbc Q{5}\or\bbb R\or\bbc S{4.2}\or\bba T{10.5}\or\bbc U{5}\or
  \bba V{12}\or\bba W{16.5}\or\bba X{11}\or\bba Y{11.7}\or\bba Z{7.5}\fi}}

\newcommand{\vs}{\vspace{0.25cm}}

\newtheorem{theorem}{Theorem}
\newtheorem{itlemma}{Lemma}[section]
\newtheorem{itproposition}[itlemma]{Proposition}
\newtheorem{itcorollary}[itlemma]{Corollary}
\newtheorem{itremark}[itlemma]{Remark}
\newtheorem{itremarks}[itlemma]{Remarks}
\newtheorem{itdefinition}[itlemma]{Definition}
\newtheorem{itexample}[itlemma]{Example}

\newenvironment{lemma}{\begin{itlemma}\rm}{\end{itlemma}} 
\newenvironment{remark}{\begin{itremark}\rm}{\end{itremark}} 
\newenvironment{remarks}{\begin{itremarks} \rm}{\end{itremarks}}
\newenvironment{corollary}{\begin{itcorollary}\rm}{\end{itcorollary}}
\newenvironment{proposition}{\begin{itproposition}\rm}{\end{itproposition}}
\newenvironment{definition}{\begin{itdefinition}\rm}{\end{itdefinition}}
\newenvironment{example}{\begin{itexample}\rm}{\end{itexample}}
\newenvironment{fact}{\noindent {{\bf Fact}}:\ \ }{\hfill \medskip}
\newenvironment{claim}{\noindent {\em Claim}. \ \ }{\hfill \medskip}
\newcommand{\be}[1]{\begin{equation}\label{#1}}
\newcommand{\ee}{\end{equation}}
\newcommand{\bl}[1]{\begin{lemma}\label{#1}}
\newcommand{\br}[1]{\begin{remark}\label{#1}}
\newcommand{\brs}[1]{\begin{remarks}\label{#1}}
\newcommand{\bt}[1]{\begin{theorem}\label{#1}}
\newcommand{\bd}[1]{\begin{definition}\label{#1}}
\newcommand{\bp}[1]{\begin{proposition}\label{#1}}
\newcommand{\bc}[1]{\begin{corollary}\label{#1}}
\newcommand{\bfact}[1]{\begin{fact}\label{#1}}
\newcommand{\bex}[1]{\begin{example}\label{#1}}
\newcommand{\ec}{\end{corollary}}
\newcommand{\efact}{\end{fact}}
\newcommand{\eex}{\end{example}}
\newcommand{\el}{\end{lemma}}
\newcommand{\er}{\end{remark}}
\newcommand{\ers}{\end{remarks}}
\newcommand{\et}{\end{theorem}}
\newcommand{\ed}{\end{definition}}
\newcommand{\ep}{\end{proposition}}
\newcommand{\epr}{\end{proof}}
\newcommand{\bpr}{\begin{proof}}
\newcommand{\bcl}{\begin{claim}}
\newcommand{\ecl}{\end{claim}}

\newcommand{\bi}{\begin{itemize}}
\newcommand{\ei}{\end{itemize}}
\newcommand{\ben}{\begin{enumerate}}
\newcommand{\een}{\end{enumerate}}

\usepackage{graphicx}

\title{Dynamical Decomposition of Bilinear Control Systems subject to Symmetries}
\author{Domenico D'Alessandro \and Jonas T. Hartwig}
\date{\today}

\begin{document}

\maketitle 
\begin{abstract}
We describe a method to analyze and decompose the dynamics of a bilinear control system  subject to symmetries.  The method is based on the concept of {\em generalized Young symmetrizers}  of representation theory. It naturally applies to the situation where the system evolves on a  tensor product space and there exists a finite group of symmetries for the dynamics which interchanges the various factors. This is the case  for  {\it quantum mechanical multipartite} systems, such as spin networks, where each factor of the tensor product represents the state of one of the component systems. { We present several examples of application.}  
\end{abstract}

\vs 

\vs
\vs

\vs 

\vs
\vs

\section{Introduction}
In geometric control theory, one often considers bilinear systems of the form 
\be{Sys}
\dot X=AX+\sum_{j=1}^m B_j u_j X, \qquad X(0)={\bf 1}, 
\ee
where $X$ varies in a matrix Lie group and $A$ and $B_j$'s belong to 
 the corresponding Lie algebra, with $u_j$ the controls, and ${\bf 1}$ the identity of the group. It is a well known fact  \cite{JS} that the { {\it reachable set}} for (\ref{Sys}) 
 is the connected  
 Lie group $e^{\cal L}$, containing the identity ${\bf 1}$, corresponding 
 to the Lie algebra ${\cal L}$ generated by $A$ and $B_j$'s, assuming that $e^{\cal L}$ is compact.  
 Therefore system (\ref{Sys}) is called {\it controllable} if $e^{\cal L}$ is some `natural' Lie group where the system is supposed to evolve. Common examples are the special orthogonal 
 group $SO(N)$ and  the unitary group $U(N)$ which 
 appears in applications of control theory to quantum mechanics. If the system of interest has the form 
\be{SysVec}
\dot \psi=A\psi +\sum_{j=1}^m B_j u_j \psi, \qquad \psi(0)=\psi_0,  
\ee   
where $\psi$ belongs to a {\it vector space} $\tilde V$,  the 
reachable set from $\psi_0$ is $\{X \psi_0 \, | \, X \in e^{\cal L}  \} $. This fact has had many applications. In particular,  for  controlled {\it quantum mechanical systems}, in finite dimensions,  the equation (\ref{Sys})-(\ref{SysVec}) is the {\it Schr\"odinger equation} incorporating  a semiclassical control field $\vec u(t):=(u_1,...,u_m)$) (see, e.g., \cite{Mikobook} for examples of modeling). In this case, the matrices $A$ and $B_j$ in (\ref{Sys}), (\ref{SysVec}) belong to the Lie algebra $u(N)$ of skew-Hermitian, $N \times N$, matrices, so that ${\cal L}$ is a Lie {\it subalgebra} of $u(N)$. The matrix $X$ in (\ref{Sys}) is called the (quantum mechanical) {\it evolution operator} and $\psi$ is the {\it state} of the quantum system belonging to a Hilbert space $\tilde V$. In this case,   controllability is said to be verified if $e^{\cal L}$ is the full unitary ($U(N)$) or special unitary ($SU(N)$) Lie group.

Although controllability is a generic property (see, e.g., \cite{Altafini}, \cite{Lloyd}), often, in 
reality, symmetries of the physical system  {and a too small number of control functions as compared to the dimension of the system} cause 
the {\it dynamical Lie algebra} ${\cal L}$, generated by $A$ and $B_j$'s, to be only  a {\it proper}
 Lie subalgebra of the natural Lie algebra associated to the model  (for example $u(N)$). The problem therefore arises to analyze the structure of this Lie algebra and to understand how this impacts the dynamics of the system  (\ref{Sys})-(\ref{SysVec}).

 {In the context of control of quantum systems, which is the main area of application we have in mind, this problem has been tackled in several references with tools of Lie algebras and representation theory (see, e.g., \cite{Polack}, \cite{Zeier}, \cite{Zimboras}). {
 One sees the vector space $\tilde V$ where $\psi$ in (\ref{SysVec}) lives as the space associated to a {\it representation} (see basic definitions of representation theory in the next section) of the Lie group $e^{\cal L}$ or the Lie algebra ${\cal L}$.}   In the paper \cite{Mikodeco}, one assumes to have a basis of the dynamical Lie algebra ${\cal L}$. Algorithms are given to decompose such a Lie algebra into Abelian and simple ideals which are its elementary components (Lie sub-algebras). Such algorithms are, for the most part, simplified and adapted versions of general algorithms presented for Lie algebras over arbitrary fields in the book \cite{DeGraaf}. The paper \cite{Polack}  identifies {\it two} causes of uncontrollability for quantum systems. On one hand, the presence of symmetries, i.e., operators commuting with the full dynamical Lie algebra ${\cal L}$, implies that the given representation of ${\cal L}$ is not {\it irreducible}, that is, the  vector space $\tilde V$, where  $\psi$ in (\ref{SysVec}) lives,  {\it splits} into a number of invariant subspaces each carrying an irreducible representation of the dynamical Lie algebra ${\cal L}$. Transitions from one subspace to the other are forbidden for the dynamics which results in   uncontrollability. The second cause of uncontrollability is the fact that, even within the invariant subspaces,  the system might be not controllable because of lack of control power. In fact, the paper \cite{Polack} presents a list of possible Lie subalgebra that might appear as irreducible restrictions of ${\cal L}$ to invariant subspaces. In view of the recalled decomposition of the dynamical  Lie algebra into irreducible components, a new, weaker, notion of controllability was introduced for quantum systems called {\it subspace controllability}. This is verified when the dynamical Lie algebra is such as to act as $u(n)$ on all or some of the invariant subspaces. Subspace controllability was recently investigated for a number of quantum control systems, most notably networks of spins \cite{Xinhua}, \cite{Wang1}, \cite{Wang2}. It was shown \cite{Wang1} that, in some cases, the dimension of the largest invariant subspace grows exponentially with the number of particles in the network so, subspace controllability gives the opportunity of doing universal quantum computation on a restricted subspace even in the absence of full controllability.

From a practical point of view, for a quantum control system with a group of symmetries $G$, the question arises of {\it how} to obtain the decomposition of the dynamics into invariant subspaces.    This is the 
topic  of this paper. We focus on a specific method to obtain this which exploits the {\it duality} between representations of ${\cal L}$ and representations of $G$ (this is some times referred to as Schur-Weyl duality (cf. e.g., \cite{Goodman})). However, in this introduction, we next describe a general different method and then we discuss the drawbacks of this method to motivate instead the treatment of the rest of the paper.

  Given the dynamical Lie algebra ${\cal L} \subseteq u(N)$, one calculates a basis of the {\it commutant} of ${\cal L}$ in $u(N)$, i.e., the subspace of $u(N)$ of elements that commute with ${\cal L}$. This amounts to the solution of a system of linear equations. Being a subalgebra of $u(N)$ the commutant is a {\it reductive} Lie algebra (see, e.g., \cite{Mikodeco}), that is, it is the direct sum (i.e., vector space sum of commuting subspaces) of an Abelian subalgebra and a semisimple one. As such,  it admits a {\it Cartan subalgebra} which  is a {\it maximal Abelian subalgebra} and can be calculated with, for example, the algorithms of \cite{Mikodeco}, \cite{DeGraaf}. Elements of a basis of such a Cartan subalgebra can be simultaneously diagonalized and therefore a basis can be found so that they can be written as 
$\texttt{diag}( i{\bf 1},{\bf 0},...,{\bf 0})$, $\texttt{diag}( {\bf 0},i{\bf 1},...,{\bf 0})$,...,
$\texttt{diag}( {\bf 0},{\bf 0},...,i{\bf 1})$, for appropriate dimensions of the zero matrices ${\bf 0}$ and the identity matrices ${\bf 1}$. This basis, gives the sought for change of coordinates that transforms the Lie algebra ${\cal L}$ in block diagonal form, so that every block corresponds to an irreducible representation of ${\cal L}$. In fact,  having to commute with the above matrices, the matrices of ${\cal L}$ take a block diagonal form. Moreover,  each block corresponds to an irreducible representation of ${\cal L}$. To see this,  let $N_1$ be the dimension of such a block, and assume without loss of generality that it is the first block. If this was not irreducible, there would be another  block diagonal matrix $A$ in $u(N)$, which,  in appropriate coordinates, would have  all blocks equal to zero and the first block equal to $\texttt{diag}(-ia{\bf 1}_{d_1}, -ib {\bf 1}_{d_2})$ for $a \not=b$ and appropriate dimensions $d_1$ and $d_2$ of the identity blocks. The matrix $A$ would be commuting with all the matrices in the dynamical Lie algebra ${\cal L}$ and would be also commuting with all the matrices in the above Cartan subalgebra of the commutant. However this contradicts the fact that the Cartan subalgebra is {\it maximal} Abelian.

The above method always gives a basis such that the dynamical Lie algebra ${\cal L}$ is decomposed into its irreducible components. However it requires the explicit solutions of linear systems of equations for matrices of possibly high dimension. For example, in the case of a network of $n$ spin $\frac{1}{2}$ particles, the dimension of the state space increases as $2^n$ and therefore the above computations involve matrices in $u(2^n)$, a space of dimension $4^n$. Moreover the role of the group of symmetries $G$ is hidden  when we transform the problem into a (high dimensional) linear algebra problem. For example, if the system is a network of spin $\frac{1}{2}$'s and the symmetry group is some subgroup of the symmetric group (the permutations which leave the matrices  appearing 
in (\ref{Sys}) (\ref{SysVec}) unchanged)  such a symmetry group is suggested by the topology of the network.

This paper is devoted to presenting  an alternative to the above approach based  to the study of the representation theory of the symmetry group $G$ itself. The representation theory of finite groups is a topic for which much is known  (see, e.g., \cite{Dixon}, \cite{FH}, \cite{Goodman}, \cite{Isaacs}, \cite{Rotman}, \cite{Sengupta}, \cite{Tung}, \cite{Woit}). From the knowledge of the representations of the group of symmetries $G$  one obtains the  change of coordinates which places the Lie subalgebra of all elements of $u(N)$ which commute with $G$, $u(N)^G$ in a block diagonal form, where each block corresponds to an irreducible representation. Since the dynamical Lie algebra ${\cal L}$ is a Lie  subalgebra of $u(N)^G$, it will also be placed in the same block diagonal form. 

{
This paper is a {\it survey paper} or, perhaps more appropriately, an {\it application paper} aimed at presenting 
known results in representation theory in a self-contained fashion so that they can be used by control theorists dealing with systems of the form (\ref{Sys})-(\ref{SysVec}), and in particular for quantum systems.} 
 }

The paper is organized as follows: In section \ref{Back}, we give some background notions from representation theory including the definition and properties of {\it Generalized Young Symmetrizers} (\textit{GYS}), which play a crucial role in the method described. The method for dynamical 
decomposition is described in section  \ref{basictheo}.  It  requires identifying certain GYS's  and, in section \ref{DGYS},  we discuss how these are obtained in two special cases:  the case of the full symmetric group $S_n$ and the case of  Abelian groups. In section \ref{esempi},  we present two examples of applications to spin networks where we use the above techniques to obtain the GYS's and the decomposition. These results, in particular extend the results of \cite{AD} for {\it fully symmetric} spin networks to the case of an arbitrary number $n$ of spins, with  the computations for the case $n=4$  presented in detail.

\section{Background and Statement of the Problem}\label{Back}

\subsection{Representation theory and statement of the problem}
{ 
We shall be interested in representations, $(\pi, \tilde V)$, of groups, $G$, algebras, ${\cal A}$,  or Lie algebras, ${\cal R},$  on a finite dimensional complex inner product  space $\tilde V$ of dimensions $N$ which we can identify with $\CC^N$. The space $\tilde V$ is often called a {\it $G$-module} (or {\it ${\cal A}$-module}, or an {\it ${\cal R}$-module}).  Representations are group, algebra or Lie algebra homomorphisms  from $G$, ${\cal A}$ or ${\cal R}$ to $\texttt{End}(\tilde V)$ the space of endomorphisms on $\tilde V$, which if $\tilde V \simeq \CC^N$ can be identified with the space of $N \times N$ matrices with complex entries. Given representations of $G$, ${\cal A}$ and  ${\cal R}$, on the same space $\tilde V$,  we shall denote by ${\cal A}^G$ or  ${\cal R}^G$ the (Lie)  subalgebra of elements in ${\cal A}$ or ${\cal R}$, or more precisely of their representation, which  commute with the representation of $G$. For example, for a quantum control system (\ref{Sys}) (\ref{SysVec}), we are given a representation of the dynamical Lie algebra ${\cal L}$ generated by the "Hamiltonians" $A$ and $B_j$'s which is a subalgebra of $u(N)$ and a representation on the same space of a group of symmetries which commute with the elements of ${\cal L}$. Therefore ${\cal L} \subseteq u(N)^{G}$ the subalgebra of $u(N)$ commuting with $G$.

We fix some notations.  We shall denote by $\texttt{End}_G(\tilde V)$ the space of all endomorphisms of $\tilde V$ commuting with $G$. Given two representations $(\pi_1, \tilde V_1)$ and $(\pi_2, \tilde V_2)$, $\texttt{Hom} (\tilde V_1, \tilde V_2)$ denotes the space of homomorphisms $\phi \, : \, \tilde V_1 \rightarrow \tilde V_2$, 
$\texttt{Hom}_G (\tilde V_1, \tilde V_2)$  denotes the subspace of $\texttt{Hom} (\tilde V_1, \tilde V_2)$ of elements $\phi \in \texttt{Hom} (\tilde V_1, \tilde V_2)$ such that $\phi \pi_1(g)=\pi_2(g) \phi$ for every $g \in G$. Such a type of maps is called a {\it $G$-map}. { Analogously one can consider ${\cal A}$-maps and ${\cal R}$-maps, for algebras (${\cal A}$) and Lie algebras (${\cal R}$) representations.} If the two representations coincide $\texttt{Hom}_G (\tilde V, \tilde V)$ coincides with $\texttt{End}_G(\tilde V)$. Two representations 
$(\pi_1, \tilde V_1)$ and $(\pi_2, \tilde V_2)$ are called {\it $G$-isomorphic}  if there exists an element in $\texttt{Hom}_G (\tilde V_1, \tilde V_2)$, i.e., a $G$-map, which is also an isomorphism, a {\it $G$-isomorphism}.

Representations of groups  
are called {\it unitary} if their images are unitary matrices. Representations of Lie algebras are called unitary if their images are skew-Hermitian matrices. A representation $(\pi, \tilde V)$ is  called 
{\em reducible} if there exists a proper nonzero subspace of $\tilde V$ which is invariant under the representation, {\it irreducible} if there is no such subspace. Representations $(\pi, \tilde V)$, of finite groups as well as those of unitary groups or Lie algebras, are {\it completely reducible}, i.e., they can be decomposed into the direct sum of irreducible representations (see, e.g., \cite{FH}, \cite{Woit}). In these cases, $\tilde V$ is the direct sum of invariant subspaces for $\pi$, so that the restriction of $\pi$ to each invariant subspace is an irreducible representation.    In this case,  in appropriate coordinates,  the matrices  
$\pi(x)$, for $x$ element in the group, algebra, or Lie algebra, take a block diagonal form. The finite group case and the case of unitary representations are the cases that will be of interest for us in this paper.

In view of these notions, the problem to be solved in this paper, that we have outlined in the introduction,  is as follows:

\subsubsection*{Problem:} 
{\it Given a unitary representation of a Lie algebra ${\cal R}$,  and a unitary representation of a finite symmetry group $G$, on a finite dimensional Hilbert space $\tilde V$, find a decomposition of 
${\cal R}^G$ into its irreducible components and the associated change of coordinates in $\tilde V$. 
 }

In the case of quantum control, the Lie algebra ${\cal R}$ is $u(N)$ and if the dynamical Lie algebra ${\cal L} \subseteq u(N)$ commutes with a group of symmetries $G$, then we look for a decomposition in  irreducible representations of $u(N)^{G}$ since ${\cal L} \subseteq u(N)^{G}$. In the coordinates we find, ${\cal L}$ also takes a block diagonal form.

\vs 
{
A fundamental tool in representation theory is the following {\it Schur's Lemma} (see, e.g., \cite{Woit}, Section 2.1). 

\bt{Schur} (Schur's Lemma) Let $B$, be a group or an algebra or a Lie algebra. 
\begin{enumerate}

\item If a complex representation $(\pi, \tilde V)$ of $B$ 
is irreducible, all $B$-maps $\tilde V \rightarrow \tilde V$ are multiples  of the identity map.

\item Two irreducible representations $(\pi_1, \tilde V_1)$ and  $(\pi_2, \tilde V_2)$ are such that the space of $B$-maps is either $1-$dimensional or $0$-dimensional according to whether the two representations are isomorphic or not. 

\end{enumerate}

\et 

\bpr The two statements are  equivalent. If statement 2 holds, than taking  $(\pi_1, \tilde V_1)=(\pi_2, \tilde V_2)=(\pi, \tilde V)$ and noticing that the identity map is a $B$-map, we obtain  statement 1. Now we prove first statement 1 and then show that statement 2 follows from it. 

 For any $B$-map $\phi$  between two representations  $(\pi_1, \tilde V_1)$ and  $(\pi_2, \tilde V_2)$, the  Kernel of $\phi$ and the Image of $\phi$ are invariant subspaces for  the representations  $(\pi_1, \tilde V_1)$ and  $(\pi_2, \tilde V_2)$, respectively. Consider now a $B$-map, $\phi$ for the representation $(\pi, \tilde V)$ and let $\alpha$ be an eigenvalue of $\phi$. Then, if ${\bf 1}$ is the identity map, $\hat \phi_\alpha:=\phi-\alpha {\bf 1}$ is a $B$-map as well, and the Kernel of $\hat \phi_\alpha$ is not zero. Since 
$(\pi, \tilde V)$ is irreducible,  the Kernel must be all of $\tilde V$, that is  $\phi-\alpha {\bf 1}=0$, which proves the first statement. 
 
As for the second statement,  assume 
$\phi: \, \tilde V_1 \rightarrow \tilde V_2$ is a $B$-map.  Then because of irreducibility $Ker(\phi)=0$ or $Ker(\phi)=\tilde V_1$ and $Im(\phi)=0$ or $Im(\phi)=\tilde V_2$. If $Ker(\phi)=0$ and $Im(\phi)=\tilde V_2$ then $\phi$ is an isomorphism. In all other cases it is zero. If $\phi$ and $\gamma$ are two isomorphisms, from 
$\phi \pi_1=\pi_2 \phi$ and $\gamma \pi_1=\pi_2 \gamma$, we obtain 
$\phi \gamma^{-1} \pi_2=\pi_2 \phi \gamma^{-1}$ which using the first statement implies that $\phi$ is a multiple of $\gamma$, which proves the second statement. 

\epr 

We remark that Schur's Lemma applies to both real and complex (Lie) algebras as long as the considered representations  are complex, i.e., $\tilde V$, (or $\tilde V_{1,2}$) are complex vector spaces. We need, 
in fact, the underlying field to be algebraically closed in order to be able to always find an eigenvalue for the $B$-map of part 1. More general and abstract formulations of Schur's Lemma exist (see, e.g., \cite{Goodman} and references therein).

}

\subsection{Group algebra, regular representation and Generalized Young Symmetrizers}

Given a finite group $G$, 
the group algebra $\CC[G]:=\bigoplus_{\Pi\in G} \CC\Pi$ is the complex vector 
space with basis given by the elements of $G$ equipped with multiplication given by bilinearly extending the group operation. For example, for $G=S_3$, the symmetric group of three elements, 
 \[(12)\cdot \big(\lambda \cdot (1)+\mu\cdot (13)\big) = \lambda \cdot (12)\cdot(1)+\mu\cdot (12)\cdot(13) = \lambda  \cdot (12)+\mu \cdot (132),\] for $\lambda ,\mu\in\CC$ (here and in the following we use the convention of multiplying permutations from right to left, as compositions of transformations). If 
$\tilde V$ is a $G$-module  then it is also $\CC[G]$-module where  $\CC[G]$ acts on $\tilde V$ by linearly extending the  action of the group $G$. If we take as $\tilde V$ exactly $\CC[G]$,  the action of $G$ on $\tilde V$ gives a representation of $G$ called the {\it regular representation}. The regular representation is, in general, {\it not} irreducible and it contains, as irreducible components,  all the irreducible representations of the finite group $G$. More precisely, the following fundamental fact holds (cf., e.g., \cite{FH}): 
\bt{basicfact5} Every irreducible representation of a finite group $G$ on a vector space $\tilde V$ is $G$-isomorphic to one irreducible representation contained in the regular representation.  
\et 
{Irreducible representations may be contained (up to $G-$isomorphism) more than once in $\CC[G]$.} Their multiplicity  { is} equal to the dimension of the representation. That is, { we have (cf., e.g., \cite{FH} \S \, 3.4)
\be{dimensions}
\CC[G]=\bigoplus_j ({\cal C}_j)^{\oplus \dim {\cal C}_j},   
\ee
for the irreducible representations ${\cal C}_j\subseteq \CC[G] $ of $G$
which, in particular, implies that 
\be{sumdim}
\sum_j (\dim  {\cal C}_j)^2=\dim \CC[G]=|G|,
\ee  
the number of elements in the group $G$.  
}

\bd{YS} ({\bf Generalized  Young Symmetrizers (\textit{GYS})})  Given a finite group 
$G$, a {\it complete set of Generalized Young Symmetrizers} is a set of elements $\{P_j\}$, $j=1,\ldots,m$, of the associated group algebra  $\CC[G]$ satisfying the following properties: 
\begin{enumerate}
\item ({\it Completeness})
\be{P1}
{\bf 1} =\sum_{j=1}^m P_j;
\ee 
where ${\bf 1}$ is the identity of the group.
\item ({\it Orthogonality}) 
\be{P2}
P_jP_k=\delta_{j,k}P_j , \quad \forall j,k;
\ee
where $\delta_{j,k}$ is the Kronecker delta. 
\item ({\it Primitivity}) 
For every $g \in G$ 
\be{P3}
P_jgP_j=\lambda_g P_j,  
\ee
for every $P_j$ with  $\lambda_g$ a scalar  that depends on $g$.  
\end{enumerate}
\ed
Generalized Young symmetrizers are called a {\it complete set of primitive orthogonal idempotents} in ring theory. Their significance in representation theory is that they generate left  ideals 
in the group algebra $\CC[G]$ which correspond to irreducible sub-representations of 
the regular representation of $G$. In particular given a set of GYS's,  we can write $\CC[G]$ as 
\be{sumofideals}
\CC[G]=\CC[G]{\bf 1}= \CC[G] (\sum_j P_j)={\cal C}_1+{\cal C}_2+\cdots + {\cal C}_q, 
\ee
where ${\cal C}_j:=\CC[G] P_j$, $j=1,\ldots,q$,  is a left ideal of $\CC[G]$ and, in particular, an invariant subspace of  $G$ in $\CC[G]$, i.e., a sub-representation of the regular representation. { Fix $j \geq 2$ and let $x \in {\cal C}_j \cap 
{\cal C}_1+   {\cal C}_2+\cdots + {\cal C}_{j-1}$. Then there 
exist $A_1,A_2,...,A_j$ in $\CC[G]$ such that $x=A_jP_j=A_1 P_1+A_2 P_2+\cdots + A_{j-1}P_{j-1}$. Multiplying on the right  by $P_j$ and using (\ref{P2}) we obtain $x=0$.} Therefore the sum in (\ref{sumofideals}) is a {\it direct} sum of sub-modules, i.e., 
$\CC[G]=\bigoplus_{j=1}^q{\cal C}_j$. According to Theorem III.3 of the Appendix III of \cite{Tung}, condition (\ref{P3}) is necessary and sufficient so that 
the ideal ${\cal C}_j$ is {\it minimal} which means that it does not properly 
contain any other ideal. This is usually expressed by saying that the 
idempotent $P_j$ is {\it primitive} and in terms of representations it means that the {\it representation associated with ${\cal C}_j$ is  irreducible}. { Furthermore, according to Theorem III.2 in Appendix III of \cite{Tung}, GYS's, $P_j$, always exist, so that the irreducible sub-modules ${\cal C}_j$ of 
$\CC[G]$ can always be written as ${\cal C}_j=\CC[G]P_j$.} 

Primitive, orthogonal idempotents are called {\it Young Symmetrizers} in the context of the symmetric group $S_n$ and therefore we use here the terminology `Generalized Young Symmetrizers' to refer to the case of a general finite group.  In the case of the symmetric group,  Young symmetrizers are obtained from Young tableaux as summarized in many textbooks such as \cite{FH}, \cite{Goodman}, and \cite{Tung}. We shall review the main points in subsection \ref{DGYS}.

{
Another property of GYS's which we shall require is of being {\it Hermitian}. To define this property define a conjugate-linear map on $\CC[G]$, which is denoted by $^\dagger$ and it is defined on elements of $G$, like $g^{\dagger}:=g^{-1}$ and extended to  $\CC[G]$ by conjugate linearity, that is, 
$\left( \sum_j a_j g_j \right)^\dagger  = \sum_j \bar a_j g_j^\dagger$, for $g_j \in G$ and $a_j \in \CC$. With this definition we may require that the GYS's are Hermitian, i.e.,    
\be{P4}
P_j=P_j^\dagger, \qquad j=1,2,...,q.
\ee

In our context, we have a $G$-module, $\tilde V$, which is extended by linearity to be a $\CC[G]$-module.  We shall  see elements in the group algebra $\CC[G]$  as operators on the  vector space $\tilde V$.  We can view, in particular any   GYS  $\{P_j\}$ as an operator on $\tilde V$.  For $a:=\sum_j a_j g_j$, 
we have $\pi(a)=\sum_j a_j \pi(g_j)$ and $\pi(a^{\dagger})=\sum_j \bar a_j \pi(g_j^{\dagger})=\sum_j \bar a_j (\pi(g_j))^{-1}$. 
If the representation $(\pi,G)$ is unitary $(\pi(g_j))^{-1}=(\pi(g_j))^{\dagger}$ so that 
$\pi(a^\dagger)=\sum_j \bar a_j (\pi(g_j))^\dagger$, so that $\pi(a^\dagger)=(\pi(a))^\dagger$. So if $a$ is Hermitian ($a=a^\dagger$), its image under a unitary representation will also be Hermitian in the standard sense of Hermitian matrices. 

}

{The Hermiticity  property will be important  in our treatment of representations of Lie subalgebras of $u(N)$, in applications to quantum mechanical systems.  We will take advantage of recent results of \cite{Her1} and \cite{KS} which show how to modify the standard procedure to obtain Young Symmetrizers in order to obtain {\it Hermitian} Young Symmetrizers, for the case of the symmetric group.}


{ 
\section{Decomposition of the Dynamics}\label{basictheo}
We now assume that, for a group $G$, we have a complete set of  GYS's. We show how this information 
can be used to decompose a Lie algebra ${\cal R}^G$, i.e., the subalgebra of a Lie algebra ${\cal R}$ consisting of all elements of ${\cal R}$ commuting with $G$. This gives the decomposition of the dynamics induced by the symmetries in $G$, and in the associated coordinates, the system (\ref{SysVec}) (and (\ref{Sys})) can be put in a block diagonal form. We shall discuss in the following section how GYS's can, in certain cases,  be obtained. 
}

{
When we are given a system (\ref{SysVec}) with $\psi$ varying in a complex vector space $\tilde V$, the space $\tilde V$ simultaneously carries representations of the dynamical Lie algebra ${\cal L}$, a natural Lie algebra ${\cal R}$ (for example $u(N)$), with ${\cal L}\subseteq {\cal R}$,  a finite group of symmetries $G$, its group algebra $\CC[G]$, as well as $\texttt{End}(\tilde V)$ and $\texttt{End}_G(\tilde V)$. We are ultimately interested in ${\cal R}^G$, since ${\cal L} \subseteq  {\cal R}^G$, but we describe the representation of $\texttt{End}_G(\tilde V)$ first.  Since ${\cal R}^G={\cal R} \cap \texttt{End}_G(\tilde V)$, the representation of ${\cal R}^G$ is obtained by restricting the elements of the representation of $\texttt{End}_G(\tilde V)$ to the ones that also belong to the representation of ${\cal R}$ (for example skew-Hermitian matrices if ${\cal R}=u(N)$). Given the complete set of GYS's, $\{P_j\}$ and their representations (as elements of the group algebra $\CC[G]$), which, with some abuse of notation,  we still denote by $\{P_j\}$,  we consider a decomposition of $\tilde V$ as 
\be{tildeVdeco}
\tilde V=\oplus_j^q P_j \tilde V. 
\ee
To see that this holds, first notice that for ever $y \in \tilde V$, $y=(\sum_j P_j)y=\sum_j P_jy$, because of the completeness property (\ref{P1}). Moreover, for $j \geq 2$,  
if $x \in P_j \tilde V \cap ( P_1 \tilde V+ P_2 \tilde V+ \cdots P_{j-1} \tilde V)$, i.e., $x=P_j x_j= P_1 x_1+ P_2 x_2+ \cdots P_{j-1} x_{j-1}$, applying $P_j$ to both sides and using the orthogonality 
relation (\ref{P2}), we obtain that $x=0$, and therefore the sum (\ref{tildeVdeco}) is a direct sum 
(cf. (\ref{sumofideals})). We choose a basis of $\tilde V$ by putting together bases of $P_1 \tilde V$, 
$P_2 \tilde V$,...,$P_q \tilde V$. Furthermore, we group together bases corresponding to GYS's, $P_j$, which give isomorphic ideals, ${\cal C}_j$, in the group algebra $\CC[G]$.

We now analyze the matrix representation of elements in $\texttt{End}_G(\tilde V)$ in this basis. If 
$F \in \texttt{End}_G(\tilde V)$, then, for every $j$, $FP_j=P_jF$, and therefore $P_j \tilde V$ is invariant under $F$. This implies that, in the chosen basis,  $F$ has a block diagonal form 
\be{ablockdiag}
F=\begin{pmatrix} 
{\begin{matrix} A_1 & &  \\  
& \xddots &   \\
& & {A_{m_A}}
\end{matrix}} & \rvline &   & \rvline & & \rvline & \cr 
\hline 
& \rvline & {\begin{matrix} B_1 & &  \\  
& \xddots &   \\
& & {B_{m_B}} 
\end{matrix}} & \rvline & & \rvline &  \cr 
\hline 
& \rvline & & \rvline &  & \rvline & \cr 
& \rvline & & \rvline & \xddots & \rvline & \cr 
& \rvline & & \rvline &   & \rvline & \cr 
\hline 
& \rvline & & \rvline & & \rvline & 
{\begin{matrix} C_1 & &  \\  
& \xddots &   \\
& & {C_{m_C}}
\end{matrix}}\cr
\end{pmatrix}
\ee   
where we denoted with the same letter blocks corresponding to isomorphic ideals in the group algebra. We remark that, depending on the representation at hand, some ideals ${\cal C}_j$ and corresponding GYS's $P_j$ might not be present in the above decomposition meaning that some $P_j\tilde V$, might be zero.

Now,  we want to obtain more information on the nature of the submatrices in $F$ in (\ref{ablockdiag}) and we want  to study the form of the representation of $G$ in the same basis. From this, the {\it duality} between the representations of $G$ and $\texttt{End}_G(\tilde V)$, will be apparent. This will rely on the following three propositions whose proofs are presented in the next subsection.  

\bp{fromTung} Two left ideals in $\CC[G]$, ${\cal C}_1$  and ${\cal C}_2$,  generated by GYS's $P_1$ and $P_2$, respectively, are $G$-isomorphic if and only if there exists an $r \in G$ such that 
\be{rnot0}
P_1rP_2\not=0.
\ee
\ep

\bp{irreducibil9} For each GYS, $P_j$, $P_j \tilde V$ is either zero or it is an irreducible 
representation of $\texttt{End}_G(\tilde V)$.  
\ep

\bp{irre9} Consider two nonzero $\texttt{End}_G(\tilde V)$-modules $P_j \tilde V$ and $P_k \tilde V$. They 
are   $\texttt{End}_G(\tilde V)$-isomorphic if and only if ${\cal C}_j$ and ${\cal C}_k$ are isomorphic as $G$-modules. In this case, an $\texttt{End}_G(\tilde V)$-isomorphism,  
$P_j \tilde V \rightarrow  P_k \tilde V$, is given by $P_k r P_j$, for any $r \in G$ such that $P_k r P_j \not=0$, which exists because of Proposition \ref{fromTung}. 
\ep       
Before proving these facts, we see how they impact the form of the representation of $\texttt{End}_G(\tilde V)$ in (\ref{ablockdiag}). Sub-blocks of the matrix $F$  corresponding to isomorphic ideals ${\cal C}_j$, must have the same dimension, according to Proposition \ref{irre9}. Therefore the blocks $A_1,...,A_{m_A}$ have all the same dimension in (\ref{ablockdiag}), and the same is true for $B_1,...,B_{m_B}$, and so on.   Moreover, we can refine our choice of the basis as follows. Let $P_{j_1}\tilde V$,...,$P_{j_m}\tilde V$ be a maximal set of subspaces in the decomposition 
isomorphic to each-other. Choose a basis of for $P_{j_1} \tilde V$, $\{x_1,...,x_{d_1}\}$, and using the isomorphism of Proposition \ref{irre9}, choose a basis for $P_{j_2} \tilde V$, given by  
$\{P_{j_2}r P_{j_1}x_1,...,P_{j_2}r P_{j_1}x_{d_1}\}$, for $r \in G$ such that   $P_{j_2}r P_{j_1} \not=0$. If, for $b=1,...,d_1$, $Fx_b=\sum_{l=1}^{d_1} a_{lb}x_l$ for some coefficients $a_{lb}$, then 
$$
F(P_{j_2}r P_{j_1}x_b)=P_{j_2}r P_{j_1} Fx_b=P_{j_2}r P_{j_1} \sum_{l=1}^{d_1} a_{lb}x_l=
\sum_{l=1}^{d_1} a_{lb}(P_{j_2}r P_{j_1}x_l). 
$$ 
Therefore, the coefficients of the matrix corresponding to $F$, $\{ a_{lb} \}$ are the same for the actions on $P_{j_1} \tilde V$ and $P_{j_2} \tilde V$, and the matrix representations are the same. We can repeat this argument for the remaining $P_{j_3}\tilde V$,...,$P_{j_m}\tilde V$, if any, and show that all the matrices $A_1,$...,$A_{m_A}$, in (\ref{ablockdiag}) are equal, i.e., $A_1=A_2=\cdots=A_{m_A}=A$. Repeating the same argument for all other  sets  of isomorphic spaces, we find that, in the given basis, the matrices of representations of $\texttt{End}_G(\tilde V)$ have the form 
\be{matfor5}
F=\begin{pmatrix} &\diagentry{{\bf 1}_{m_A} \otimes A} & \\
 & &\diagentry{{\bf 1}_{m_B} \otimes B} &\\
& &&\diagentry{\xddots} &\\
&&& &\diagentry{{\bf 1}_{m_C} \otimes C} &\\
\end{pmatrix}
\ee
where the numbers $m_A$, $m_B$,...,$m_C$ describe how many times isomorphic representations enter the given representation of $\texttt{End}_G(\tilde V)$. Moreover the matrices $A$, $B$,...,$C$ correspond to irreducible representations of   $\texttt{End}_G(\tilde V)$.

We now study the form of the representation of $G$ in the above basis. Fix one GYS, $P_1$ and let $P_2,$...,$P_m$ the GYS's corresponding to isomorphic $\texttt{End}_G(\tilde V)$-modules and isomorphic $G$-submodules in $\CC[G]$ (cf. Proposition \ref{irre9}). If $g \in G$ and $y:=P_1 x  \in P_1 \tilde V$, we have 
$$
g y=g P_1 x=  \left(\sum_{j=1}^m P_j+ \sum_{j\notin \{1,...,m\}} P_j \right) g P_1 x=
\sum_{j=1}^m P_jgP_1 x, 
$$
where we used the completeness relation (\ref{P1}) and Proposition \ref{fromTung}. This shows that $\bigoplus_{j=1}^m P_j \tilde V$ is invariant under $g$ and therefore (repeating this argument for every set of isomorphic spaces) that the  matrix corresponding to $g$ takes a block diagonal form where each block corresponds to  a (large) block in (\ref{matfor5}) and it is of dimension $m_A d_A \times m_A d_A $,  $m_B d_B \times m_B d_B $,...,$m_C d_C \times m_C d_C $. Here the integers  $m_{A,B,....,C}$ indicate how many times isomorphic subspaces $P_j \tilde V$ appear in the representation of $\texttt{End}_G(\tilde V)$ (cf. formula (\ref{ablockdiag})), the integers  $d_{A,B,...,C}$ denote their dimensions. 
Let us focus on the first large block and indicate the number of occurrences simply by  $m$ and the dimension simply by $d$. If $\vec e_1,..., \vec e_d$ is a basis of $P_1 \tilde V$, the chosen basis is   
$\vec e_1,..., \vec e_d, \Phi_{2,1}\vec e_1,...,\Phi_{2,1}\vec e_d,...., \Phi_{m,1}\vec e_1,...,\Phi_{m,1}\vec e_d$, where $\Phi_{j,1}$, $j=2,...,m$, is the $\texttt{End}_G(\tilde V)$-isomorphism, $P_1 \tilde V \rightarrow P_j \tilde V$ chosen above. Therefore the basis for $\bigoplus_{j=1}^m P_j \tilde V$, is given by $P_{j}\Phi_{j,1} \vec e_k$, $j=1,...,m$, $k=1,...,d$, ordered first by $j$ and then by $k$, where we set $\Phi_{1,1}$ equal to the identity matrix. Now, for $g \in G$, calculate $gP_j \Phi_{j,1} \vec e_k$. This gives 
$$
gP_j \Phi_{j,1} \vec e_k=\sum_{l=1}^m P_l g P_j \Phi_{j,1} \vec e_k.
$$
The element $P_l g P_j$ is either zero or, according to Proposition \ref{irre9}, is an $\texttt{End}_G(\tilde V)$-isomorphism $P_j \tilde V \rightarrow P_l \tilde V$. Therefore,  $P_l g P_j \Phi_{j,1}$ is an 
$\texttt{End}_G(\tilde V)$-isomorphism $P_1 \tilde V \rightarrow P_l \tilde V$. According to Schur's Lemma, Theorem \ref{Schur}, the space of $\texttt{End}_G(\tilde V)$-isomorphisms  $P_1 \tilde V \rightarrow P_l \tilde V$ is one-dimensional. Therefore    $P_l g P_j \Phi_{j,1}=\lambda_{l,j}(g) \Phi_{l,1}$, and we have 
$$
gP_j \Phi_{j,1} \vec e_k=\sum_{l=1}^m \lambda_{l,j}(g) \Phi_{l,1} \vec e_k,  
$$
with $\lambda_{l,j}(g)$ possibly zero for some $g \in G$. Therefore, by defining $\Lambda_m=\Lambda(g)$, the $m \times m$ matrix $\{ \lambda_{l,j} (g)\}$, the matrix corresponding to $g$ in the given basis of $\bigoplus_{l=1}^m P_l \tilde V$ has the form $\Lambda_m(g) \otimes {\bf 1}_d$. Repeating this 
for every set of isomorphic representations, we find that  the representation of $g$ on $\tilde V$ has the form 
\be{repreg}
g=\begin{pmatrix} & &\diagentry{{\Lambda}_{m_A}^A \otimes {\bf{1}}_{d_A}} & & \\
 & & &\diagentry{ {\Lambda}_{m_B}^B \otimes {\bf{1}}_{d_B}}    && \\
& & &&\diagentry{\xddots} && \\
& &&& &\diagentry{{\Lambda}_{m_C}^C \otimes {\bf{1}}_{d_C}} && \\
\end{pmatrix}. 
\ee
Comparing formula (\ref{repreg}) with formula (\ref{matfor5}) reveals the duality of the representations of $\texttt{End}_G(\tilde V)$ and $G$. The commutativity of the two representations is also made clear in the given basis. Moreover, the dual roles of the integers $m_{A,B,...,C}$ and $d_{A,B,...,C}$ is also apparent. In the representation of $\texttt{End}_G(\tilde V)$, $m$ is the number of isomorphic copies of a certain $P_j \tilde V$ in $\tilde V$ of dimension $d$. In the representation of $G$, the roles of $m$ and $d$ are reversed. The number $d$ represents the {\it dimension}  of a sub-representation of $G$ and $m$ represents {\it how many times it occurs}.

If the Lie algebra ${\cal R}$ for which we want to study the representation of ${\cal R}^G$ is not the full $\texttt{End}(\tilde V)$, we can take ${\cal R}^G={\cal R} \cap \texttt{End}_G(\tilde V)$ and take 
in (\ref{matfor5}) the matrices ${A,B,...,C}$, so that the full matrices give the representation 
of ${\cal R}^G$. For example,  if ${\cal R}=u(N)$ and we look for the representation of $u(N)^{G}$, then we will take matrices $A,B,...,C$ skew-Hermitian but otherwise arbitrary. In this case, it is important to point out that the GYS's have to give an orthonormal change of coordinates. This is achieved if,  in addition to properties (\ref{P1}), (\ref{P2}) and (\ref{P3}), we have the Hermitian property (\ref{P4}). The methods to find GYS described in the next section guarantee that this is the case.

\subsection{Proofs of  Propositions \ref{fromTung}, \ref{irreducibil9}, and \ref{irre9}.}

 For the proof of Proposition \ref{fromTung},  we follow \cite{Tung},
  Theorem III.4 in Appendix III. For the proofs of Propositions \ref{irreducibil9} and \ref{irre9}, we combine the treatment of \cite{Goodman} (cf. Theorem 4.2.1) which gives the results  for  $\texttt{Hom}_G({\cal C}_j, \tilde V)$ with Theorem 9.7 of  
\cite{Sengupta}  which says that $\texttt{Hom}_G({\cal C}_j, \tilde V)$ and $P_j \tilde V$ are isomorphic $\texttt{End}_G(\tilde V)$-modules.

\subsubsection{Proof of Proposition \ref{fromTung}}

\bpr 
First assume that (\ref{rnot0}) holds and consider the  $G$-map $\Phi(x):=xP_1rP_2$, 
${\cal C}_1 \rightarrow {\cal C}_2$. The fact that this is a $G$-map follows easily since, $\forall g \in G$, $\Phi(gx)=(gx)P_1rP_2=g(xP_1rP_2)=g \Phi(x)$. We remark that since $P_1rP_2\not=0$ the map 
$\Phi$ is not zero on ${\cal C}_1$. In fact, $\Phi({\cal C}_1)=\Phi({\CC[G]}P_1)={\CC[G]}P_1rP_2$, which in particular contains $P_1rP_2$. Therefore according to Schur lemma, Theorem \ref{Schur}, ${\cal C}_1$ and ${\cal C}_2$ are $G$-isomorphic.

Viceversa assume that there is a $G$-isomorphism, $\Phi:$ ${\cal C}_1 \rightarrow {\cal C}_2$. Then $\Phi(P_1)$ must be different from zero otherwise we would have $\Phi({\cal C}_1)=\Phi(\CC[G]P_1)=
\CC[G] \Phi(P_1)=0$. Moreover $\Phi(P_1)=\Phi(P_1)P_2=\Phi(P_1P_1 )P_2=P_1\Phi(P_1)P_2 \not=0$, where the first equality is due to the fact that $\Phi(P_1) \in {\cal C}_2$ and the last one to the fact that $\Phi$ is a $G$-map. Therefore since there exists an element $S$ in $\CC[G]$ ($S=\Phi(P_1)$) such 
that $P_1SP_2 \not=0$, there must exist $r \in G$ such that $P_1rP_2 \not=0$. Otherwise we would 
have $P_1SP_2 =0$ for any $S \in \CC[G]$. 

\epr        

\subsubsection{  Proof of Proposition \ref{irreducibil9}} 

\bpr

Assume $\vec x \in P_j \tilde V$ and $\vec y \in P_j \tilde V$ both different from zero (we are assuming $P_j \tilde V\not=0$). We shall find an element $R \in \texttt{End}_G(\tilde V)$ such that 
$R \vec x=\vec y$. Since $\vec x$ and $\vec y$ are arbitrary, this will imply irreducibility of the $\texttt{End}_G(\tilde V)$-module, $P_j \tilde V$. Consider ${\cal C}_j \vec x$ which is a $G$-module. The map $\Phi_x \, : \,  {\cal C}_j \rightarrow  {\cal C}_j \vec x $, given by $\Phi_x(a)=a\vec x$ is a $G$-map. Moreover it is injective since ${\cal C}_j$ is irreducible (the Kernel would be a sub-representation (cf. Theorem \ref{Schur})). Since $\Phi_x$ is surjective by definition ${\cal C}_j$ 
and  ${\cal C}_j \vec x$ are $G$-isomorphic. The same can be said for  ${\cal C}_j$ 
and  ${\cal C}_j \vec y$, with a map $\Phi_y$. We have therefore a $G$-isomorphism 
$\Phi_y \circ \Phi_x^{-1}$ from ${\cal C}_j \vec x$ to ${\cal C}_j \vec y$. In particular, 
$\Phi_x^{-1}(\vec x)=P_j$, so that $\Phi_y \circ \Phi_x^{-1}(\vec x)=P_j \vec y=\vec y$. 
Let $\Phi$ be any linear extension of $\Phi_y \circ \Phi_x^{-1}$ to $\tilde V$. The map 
\be{mappaerre}
R:=\frac{1}{|G|} \sum_{g\in G} g \Phi g^{-1}, 
\ee
is in $\texttt{End}_G(\tilde V)$ and coincides with $\Phi$ on ${\cal C}_j \vec x$. Applying $R$ to $\vec x$, we get $\vec y$. 
\epr 

\subsubsection{Proof of Proposition \ref{irre9}}

\bpr 
 First assume that ${\cal C}_j$ and ${\cal C}_k$ are $G$-isomorphic. Then, according to Proposition \ref{fromTung} there exists $r\in G$ such that $P_k r P_j \not=0$. The map $P_k r P_j$ is an $\texttt{End}_G(\tilde V)$-map and it is not zero on $P_j \tilde V$ 
(otherwise it would be zero on all of $\tilde V$ and therefore it would be zero). Because of the irreducibility of $P_j \tilde V$ and $P_k \tilde V$, from Schur's  Lemma, it follows that $P_j \tilde V$ and $P_k \tilde V$, are $\texttt{End}_G(\tilde V)$-isomorphic. 

Viceversa, assume that $P_j \tilde V$ and $P_k \tilde V$, are $\texttt{End}_G(\tilde V)$-isomorphic, and both non-zero. Let $\Psi$ be an $\texttt{End}_G(\tilde V)$-isomorphism, $\Psi \, : \, 
P_j \tilde V \rightarrow P_k \tilde V$. Assume by contradiction that  ${\cal C}_j$ and ${\cal C}_k$ 
are not $G$-isomorphic. We show that  $\Psi$ must be necessarily equal to zero, which gives the desired contradiction. Consider $\vec x \not=0 $ in $P_j\tilde V$ and the corresponding 
$\Psi(\vec x) \not=0 $ in $P_k\tilde V$. Consider the (non-zero) spaces ${\cal C}_j \vec x$ and 
${\cal C}_k \Psi(\vec x)$,  and consider the $G$-map between $G$-modules ${\cal C}_j$ and  
${\cal C}_j \vec x$, $\Phi_x$, defined by $\Phi_x(a)=a \vec x$.  Let $T:={\cal C}_j \vec x \cap {\cal C}_k \Psi(\vec x)$. The pre-image of $T$ under $\Phi_x$ is a $G$-invariant subspace of ${\cal C}_j$ and since ${\cal C}_j$ is an irreducible $G$-module it must be zero or the whole ${\cal C}_j$. It cannot be the whole ${\cal C}_j$, because that would imply (repeating the same argument for ${\cal C}_k$) that ${\cal C}_j \vec x= {\cal C}_k \Psi(\vec x)$. In particular, it would imply $P_j \vec x=aP_k \Psi(\vec x)$ with $a \in \CC[G]$. However, since ${\cal C}_j$ and ${\cal C}_k$ are assumed to be not isomorphic, from Proposition  \ref{fromTung} we obtain $P_j \vec x=P_j^2 \vec x=P_j aP_k \Psi(\vec x)=0$, which is a contradiction. Therefore the subspace $T \subseteq \tilde V$ is the direct sum, $T={\cal C}_j \vec x \oplus {\cal C}_k \Psi(\vec x)$.  Let $\Pi$ be an element in $\texttt{End}(\tilde V)$ which, when restricted to $T$ gives the projection onto ${\cal C}_k \Psi(\vec x)$.  We can define $R\in {\texttt{End}_G}(\tilde V)$, by $R:=\frac{1}{|G|}\sum_{g \in G} g \Pi g^{-1}$. The endomorphism $R$ is equal to $\Pi$ when restricted to $T$. In particular, it is zero on ${\cal C}_j \vec x$ and the identity on ${\cal C}_k \Psi(\vec x)$.  We have 
$$
0=\Psi(R\vec x )=R\Psi(\vec x)=\Psi(\vec x), 
$$
which gives the desired contradiction.   
\epr 
 }

\subsection{Examples}

{
\bex{quaternioni} Let the group  $G$ be the group $Q_8$, of {\it unit quaternions} $\{\pm 1, \pm i, \pm j ,\pm k\}$ with the standard multiplication between unit quaternions $ij=k=-ji$, $jk=i=-kj$, $ki=j=-ik$. Since it has order $8$, 
in the regular representation there are two isomorphic $2-$dimensional representations, and four non-isomorphic $1-$dimensional representations. This fact can be inferred  using the 
formula $\sum_j (\dim  {\cal C}_j)^2=\dim \CC[G]=|G|$ (cf. (\ref{dimensions})), along with the known fact that the number of non-isomorphic representations in the regular representation is equal to the number of conjugacy classes in the group (cf., e.g., \cite{Serre} Theorem 7 in Section 2.5), 
which is equal to $5$ in the case of $Q_8$. Denote by $\chi^2$ the $2-$dimensional representation and by 
$\chi^1_1,\chi^1_2 ,\chi^1_3,\chi^1_4$ the four $1-$dimensional representations in the regular representation. 
Consider now, for instance, as $\tilde V$ a $7-$dimensional space and assume that the representation of $Q_8$ on $\tilde V$ has one $2-$dimensional representation isomorphic to $\chi^2$, three isomorphic $1-$dimensional representations isomorphic to $\chi_1^1$ and two isomorphic   $1-$dimensional representations isomorphic to $\chi_3^1$. We assume therefore that, in the coordinates given by the GYS's, 
 the representation of $g \in Q_8$ is given as  
\be{IOC}
g= \begin{pmatrix}A_{2 \times 2} & 0 & 0 & 0 & 0 & 0\cr 
0 & b & 0 & 0 & 0 & 0\cr 
0 & 0 & b & 0 & 0 & 0\cr 
0 & 0 & 0 & b & 0 & 0 \cr 
0 & 0 & 0 & 0 & c & 0 \cr 
0 & 0 & 0 & 0 & 0 & c\end{pmatrix},
\ee
for scalar $b$ and $c$, and $2 \times 2$ matrix  $A_{2 \times 2}$. Comparing (\ref{IOC}) with formula (\ref{repreg}) (and (\ref{matfor5})), we see that, in this case, $m_A=2$ and $d_A=1$, $m_B=1$ and $d_B=3$, and $m_C=1$ and $d_C=2$. 
The matrices that commute with the matrices in (\ref{IOC}) have the form 
\be{lok}
F=
\begin{pmatrix} 
a & 0 & 0 & 0 \cr 
0 & a & 0 & 0 \cr 
0 & 0 & B_{3 \times 3} & 0 \cr 
0 & 0 & 0 & C_{2 \times 2}
\end{pmatrix}, 
\ee
for scalar $a$, and $3 \times 3$ general matrix $B_{3 \times 3}$, and $2 \times 2$ general matrix $C_{2 \times 2}$.
The irreducible representation $\chi^2$ has dimension $m_A=2$ but it enters one time 
($d_A=1$) the representation of $Q_8$. Dually, there are $m_A=2$ isomorphic representations of $\texttt{End}_{Q_8}(\tilde V)$ which have dimension $d_A=1$. There are $m_A=2$ GYS's corresponding to the representation 
$\chi^2$. They are given by the block diagonal matrices
$$
P_1:=\begin{pmatrix} 1 & { 0}  \cr  0  & {\bf 0}_6 \end{pmatrix}, 
\qquad P_2:=\begin{pmatrix}0 & 0 & { 0}
\cr 0 & 1 & { 0}  \cr { 0 } & { 0} & {\bf 0}_5  \end{pmatrix}. 
$$ 
The irreducible representation $\chi^1_1$ has dimension $m_B=1$ but enters three times 
($d_B=3$) the representation of $Q_8$. Dually, there is  only one ($m_B=1$) isomorphic representation of $\texttt{End}_{Q_8}(\tilde V)$ which has dimension $d_B=3$. There is only one  GYS corresponding to the representation $\chi^1_1$, which, in the chosen coordinates,  is given by 
$$
 P_3:=\begin{pmatrix}  {\bf 0}_2  & { 0} & {0}\cr { 0} & {\bf 1}_{3} & { 0} \cr 
 { 0} & { 0} & {\bf 0}_2 \end{pmatrix}.  
$$
Analogously, the irreducible representation $\chi^1_3$ has dimension $m_C=1$ but enters two times 
($d_C=2$) the representation of $Q_8$. Dually, there is  only one ($m_C=1$) isomorphic representation of $\texttt{End}_{Q_8}(\tilde V)$ which has dimension $d_C=2$. There is only one  GYS corresponding to the representation $\chi^1_3$, which, in the chosen coordinates,  is given by 
$$
P_4:=\begin{pmatrix}  {\bf 0}_5 &  {0} \cr 
{0} & {\bf 1}_2 \end{pmatrix}.
$$
\eex

In the above example, we assumed that the representation of the group is already given in the `natural' basis from which the expression of the GYS's was immediately deduced. Our goal was to illustrate the duality between the representation of the group $G$ and the representation of $\texttt{End}_G(\tilde V)$. In practice, one is given a representation of $G$, and therefore of $\CC[G]$. From the knowledge of the GYS's and from 
 their images under the given representation, one obtains the change of coordinates which transforms the dynamics in the desired form.

}

We now present a simple example of application to a quantum spin network with symmetries. More examples of applications to this type of setting will be given in section \ref{esempi}.  We recall the definition of the Pauli matrices $\sigma_{x,y,z}$  which will also be used in section \ref{esempi}. 
\be{PauliMat}
\sigma_x:=\begin{pmatrix} 0 & 1 \cr 1 & 0  \end{pmatrix}, \qquad \sigma_y:=
 \begin{pmatrix} 0 & i \cr -i & 0  \end{pmatrix}, \qquad
 \sigma_z:= \begin{pmatrix}  1 & 0 \cr 0 & -1\end{pmatrix}.
\ee
\bex{centralSpin} In recent years there has been a large interest for the controllability of {\it central spin networks} (see, e.g., \cite{Arenz}, \cite{Zimboras}), i.e., networks of spin $\frac{1}{2}$ particles where one (central) spin of a given  type is connected in various ways to spins of a different  type, which may represent a bath. The control may be local, on the central spin, or global on all the spins. One possible topology of the network, which we consider here, is a linear chain with the central spin in the middle and connected with two strings of (bath) spins, of the same length. All spins are interacting with each other  via next neighbor  interaction which we assume of the Ising type. Figure \ref{Fig1} describes the configuration of such a spin network: 
\begin{figure}[ht]
\centerline{\includegraphics[width=4in,height=1.4in]{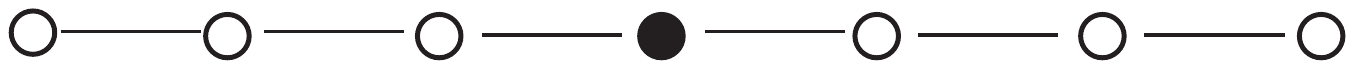}} \caption{Example of a symmetric spin network with a central spin}
\label{Fig1}
\end{figure}

Denote by $\sigma_k^j$ for $k=x,y,z$ the tensor product of $2n+1$ identities, with positions numbered from $-n$ to $n$, and with only the $j$-th position occupied  by $\sigma_k$, so that, for example, for $n=2$, $\sigma_x^1={\bf 1} \otimes {\bf 1} \otimes {\bf 1} \otimes \sigma_x \otimes {\bf 1}$. The Hamiltonians describing the 
dynamics of such a system, i.e., $A$ and $B_j$'s in (\ref{Sys}), are 
\be{newAB}
iA=\sum_{j=0}^{n-1}\sigma_z^j \sigma_z^{j+1}+ \sum_{j=0}^{-n+1}\sigma_z^j \sigma_z^{j-1}, \qquad 
iB_{x,y,z}=i \sigma_{x,y,z}^0, 
\ee
with controls $u_{x,y,z}$ representing local $x,y,z$-components of electromagnetic fields 
acting on the central spin  only.

For every $n$, such a system presents a {\it reflection} symmetry $\hat R$ since the transformation $j\leftrightarrow -j$ does not modify the Hamiltonians in (\ref{newAB}). Together with the identity, 
${\bf 1}$, $\hat R$ forms a group of symmetries for the system (\ref{Sys}),(\ref{newAB}). The two operators $P_S:=\frac{1}{2}({\bf 1}+\hat R)$, $P_A:=\frac{1}{2}({\bf 1}-\hat R)$ form a complete sets of GYS's for this group of symmetries. $P_S \tilde V$ ($P_A \tilde V$) gives all the states which are symmetric (antisymmetric) with respect to the group $\{{\bf 1}, \hat R\}$. In this basis the Hamiltonians in (\ref{newAB}) are written in block diagonal form.  

\eex

\section{Determination of the GYS's}\label{DGYS}
The above method assumes that we are able to obtain, for a given group of symmetries $G$, the corresponding (Hermitian) GYS's in the associated group algebra $\CC[G]$,  without knowing
the irreducible modules of $\CC[G]$ in advance. To the best of our knowledge, there is no general method to achieve this and it has to be done on a case by case basis. After one finds the GYS's, their image in the given representation of $G$ applied to $\tilde V$ gives the desired change of coordinates which puts the dynamics in block diagonal form.

We now discuss two cases where it  is possible to find the GYS's. In both cases, we assume that the space $\tilde V$ is the tensor product of a number $n$ of identical vector spaces $V$, i.e., $\tilde V=V^{\otimes n}$ and $G$ is a subgroup of the symmetric group $S_n$, which permutes the various factors in $V^{\otimes}$. The representation of $G$ is unitary in these cases.  The situations we shall treat are when  $G$ is the  {\it full} symmetric group,  $G=S_n$,  and when $G$ is {\it Abelian}.

\subsection{ GYS's for the symmetric group $G:=S_n$}\label{YoungTab7}
The construction of the GYS is classical  in the case where $G=S_n$ (see, e.g., \cite{Tung}) and we survey here the theory. {We shall apply it to a 
system in quantum control in the following section.} 

Conjugacy classes within $S_n$ are determined by the {\it cycle type} of a permutation, i.e., the number of cycles of a certain length. For example for $n=9$, the permutation $(123)(546)(78)(9)$ has cycle type: $2$ for cycles of length $3$, $1$ for length $2$ and $1$ for length $1$. Cycle types also correspond to {\it partitions} of $n$, i.e., sets of positive integer numbers $\lambda:= \{\lambda_1,...,\lambda_k\}$ with $\lambda_1 \geq \lambda_2 \geq \cdots \geq \lambda_k\geq 1$, and $\lambda_1+\lambda_2+\cdots \lambda_k=n$. For example, the cycle type of $(123)(546)(78)(9)$ corresponds to the partition of $n=9$, $(3,3,2,1)$ meaning that the permutations (in the given conjugacy class) have a cycle of length $3$ another cycle of length $3$, a cycle of length $2$ and a cycle of length $1$.  Partitions are encoded by {\it Young diagrams} which are diagrams composed of boxes in rows of non-decreasing lengths corresponding to the numbers in the partitions. For example, the partition of $9$, $(3,3,2,1)$ is encoded in the Young diagram   

\[  \yng(3,3,2,1) \qquad . \] 

As we have recalled in Example \ref{quaternioni}, it is a known fact in the theory of representations of finite groups that the number of non-isomorphic irreducible representations of a finite group $G$ in the regular representation is equal to the number of conjugacy classes in $G$. Therefore, in the case of the symmetric group, $S_n$, the number of irreducible representations is equal to the number of Young diagrams. In fact,  there is a stronger correspondence between Young diagrams and irreducible sub-representations of the regular representation.  If $\lambda$ is a partition of $n$, a \emph{standard Young tableaux of shape $\lambda$} is obtained from the corresponding Young diagram by distributing the numbers $1,2,\ldots,n$ over the boxes in such a way that each row and column contains  a strictly increasing sequence. For example,
\be{Tableauex}
T:=\quad \young(125,367,48,9) 
\ee
is a standard Young tableaux of shape $\lambda:=(3,3,2,1)$. The set of all standard Young tableaux of shape $\lambda$ is denoted by $\mathrm{SYT}(\lambda)$. Then there is a  correspondence between irreducible sub-representations of the regular representation, corresponding to the partition $\lambda$ (which are all isomorphic), and elements in $\mathrm{SYT}(\lambda)$. Each representation is given by { $\CC[G] P_T $} where $P_T$ is the  GYS associated to the tableaux $T$ in $\mathrm{SYT}(\lambda)$. 
The GYS $P_T$ corresponding to 
a standard Young tableaux $T$ in $\mathrm{SYT}(\lambda)$ is obtained as follows: Let $R_T$ 
be the subgroup of $S_n$ consisting of all permutations $\Pi$ which preserve the rows of $T$. Similarly, let $C_T$ be the subgroup of $S_n$ of all permutations preserving the columns of $T$. For example:

\[T:= \young(1257,36,49,8)  \qquad 
 R_T = S_{\{1,2,5,7\}}\times S_{\{3,6\}}\times S_{\{4,9\}}\quad, \qquad  
 C_T = S_{\{1,3,4,8\}}\times S_{\{2,6,9\}},  
 \]
where we omitted the singleton symmetric groups 
such as $S_{\{5\}}$ because they are the trivial group. Here, for instance, $S_{\{1,2,5,7\}}$ is the subgroup of permutations over the elements $\{1,2,5,7\}$. The \emph{row symmetrizer $r_T$ and  column anti-symmetrizer $ c_T$} 
are elements of $\CC[S_n]$ defined as follows:
\be{al}
r_T = \sum_{\sigma\in R_T} \sigma,   \qquad 
c_T= \sum_{\sigma\in C_T} (\texttt{sgn} \, (\sigma))\sigma  
\ee
The Young symmetrizer associated with $T$, $P^{'}_T$, is defined as 
$$
P^{'}_T:= r_T \cdot  c_T \qquad .  
$$
Let us consider, for example, $n=3$ and the Standard Young Tableaux  
$$
T=\young(12,3) \, .  
$$
Then $R_T=S_{\{1,2\}}$ and $C_T=S_{\{1,3\}}$ and 
$$
r_T={\bf 1}+(12), \qquad c_T={\bf 1}-(13), 
$$
$$
P^{'}_T:= r_T \cdot 
c_T=({\bf 1}+(12))({\bf 1}-(13))={\bf 1}-(13)+(12)-(12)(13)=
{\bf 1}-(13)+(12)-(132).
$$
Young symmetrizers defined this way satisfy, after being divided by a normalization factor,  the completeness property  (\ref{P1}) and the primitivity property (\ref{P3}). Therefore they give irreducible sub-representations of the regular representation. They  satisfy the orthogonality property (\ref{P2}), in general, only for small values of $n$ ($n \leq 4$). The recent paper \cite{KS}, motivated by applications in quantum chromodynamics,  shows how to modify the procedure above so that the resulting Young symmetrizers also satisfy properties (\ref{P2}) and (\ref{P4}). These recent results make the treatment in the present paper possible since we need properties (\ref{P2}) and (\ref{P4}).  In particular, property  (\ref{P4}) guarantees that the 
in the block diagonal decomposition of $u(2^n)^G$, every block is also skew-Hermitian. The procedure of \cite{KS} has been then modified in \cite{Her1} to make it significantly more efficient, in particular for large values of $n$. For our purposes however it is enough to use the original recursive algorithm of \cite{KS}.  We shall call the modified Hermitian Young Symmetrizers of \cite{KS} the {\it KS-Young symmetrizers} (from the last names of the authors of \cite{KS}). Given  a Young Tableaux $T$ corresponding to a partition of $n$, let $\texttt{Pre}(T)$ be the Young tableaux obtained from $T$ by removing the box containing the highest number and therefore corresponding to a partition of $n-1$.  For example, for the tableau $T$ in (\ref{Tableauex}), 
\be{Pretex}
\texttt{Pre}(T):=\young(125,367,48).  
\ee 
The KS-Young symmetrizer $P_T$ associated with a tableaux $T$ coincides with the standard Young symmetrizer $P^{'}_T$, if $n \leq 2$. If $n > 2$, it is obtained recursively as 
\be{recurs}
P_T= (P_{\texttt{Pre} (T)} \otimes {\bf 1}) P^{'}_T (P_{\texttt{Pre} (T)} \otimes {\bf 1}). 
\ee
It is  proved in \cite{KS} that this definition satisfies the 
requirements (\ref{P1}),(\ref{P2}), (\ref{P3}) and (\ref{P4}).

More information can be obtained from the Young tableau $T$ even without calculating the corresponding KS-Young symmetrizer $P_T$. For instance, the dimension of $\texttt{Im} (P_T)$ is equal to (cf. Lemma 3 in \cite{KS})
\be{formuladime}
\dim (\texttt{Im} P_T)=\frac{\prod_{l=1}^r \prod_{k=1}^{\lambda_l} (N-l+k)}{\texttt{Hook}(T)}.  
\ee 
Here $N=\dim(V)$,  $r$ is the number of rows of the Young tableaux, $\lambda_l$ the number of boxes in the $l$-th row, 
$\texttt{Hook}(T)$ is the {\it Hook length} of the Young diagram associated 
with $T$. It is calculated by considering, for each box of the Young diagram, the number of boxes directly to the right + the number of boxes directly below + 1 and then taking the product of all the numbers obtained. For example the Hook length of the Young tableau in (\ref{Tableauex}) is 2160. It follows from formula  (\ref{formuladime}) that if the number of rows of the tableaux is greater than the dimension $N$ of the vector space $V$, then 
$\dim (\texttt{Im} (P_T))=0$.

}

\subsection{GYS's for finite Abelian groups}

Let $G$ be a finite Abelian group. It follows from Schur's Lemma that every irreducible representation is one dimensional.\footnote{Since any element of the representation acts as a multiple of the identity, irreducibility can only occur in dimension 1.} In the following,  we shall 
use some concepts concerning the {\it character} $\chi$ of a representation $\rho$ (cf., e.g., Lecture 2 in \cite{FH}). This is a function 
$G \to \CC$ defined as $\chi(g)=\texttt{Tr} \rho(g)$, for $g \in G$. Various properties of characters of representations can be found in the representation theory texts we have cited. One property that we will use, and that directly follows from the definition,  is that the character of the direct sum of two representations is the sum of the characters (cf. Proposition 2.1 in \cite{FH}).   Characters corresponding to irreducible (and therefore one dimensional) representations are called {\it irreducible characters}. There is a one to one correspondence between irreducible characters and irreducible representations.  Every irreducible 
character is a group homomorphism $\chi:G\to\CC^\times$, whose image is contained in the unit circle $S^1$, in    
$\CC^\times$, the complex plane without the origin. Recall that from formula (\ref{dimensions}) (with $\dim({\cal C}_j)=1$) there are $|G|$ different irreducible representations in the regular representation and therefore $|G|$ different characters.  To each such character $\chi$ we associate an element $P_\chi$ of the group algebra $\CC[G]$ as follows:

\be{abelianGYS}
P_\chi := \frac{1}{|G|} \sum_{g\in G} \chi(g) g.  
\ee

\bp{GYSAb}
The set $\{P_\chi\}$ where $\chi$ ranges over the set of all possible irreducible characters, in the regular representation,  forms a complete set of Hermitian GYS for the group $G$, i.e., it satisfies properties (\ref{P1})-(\ref{P3}) and (\ref{P4}).  
\ep

\begin{proof}

Consider the following calculation.

\begin{align*}
P_\chi P_{\chi^{'}}
&= \frac{1}{|G|^2} \sum_{g,g^{'}\in G} \chi(g)\chi^{'}(g^{'})gg^{'} \\
&= \frac{1}{|G|^2} \sum_{h\in G} \big(\sum_{\substack{g,g^{'}\in G\\ gg^{'}=h}} \chi(g)\chi^{'}(g^{'})\big)h \\
&= \frac{1}{|G|} \sum_{h\in G} \big( \frac{1}{|G|} 
\sum_{g\in G} \chi(g)\chi^{'}(g^{-1})\big) \chi^{'}(h)h \\
&= \frac{1}{|G|}  \delta_{\chi \chi^{'}}  \sum_{h \in G}\chi^{'}(h) h\\
&=\delta_{\chi \chi^{'}} P_{\chi^{'}}, 
\end{align*}
where we used the character orthogonality condition 
$\frac{1}{|G|} \sum_{g\in G} \chi(g)\chi^{'}(g^{-1})=\delta_{\chi \chi^{'}}$ (cf. formula (2.10) in \cite{FH}), with the property $\bar \chi(g)=\chi(g^{-1})$.   This gives (\ref{P2}).

To see that $P_\chi$ is Hermitian, we calculate 
\[P_\chi^\dagger = \sum_{g\in G} \overline{\chi(g)} g^{-1} = \sum_{g\in G} \chi(g^{-1})g^{-1} = \sum_{h\in G} \chi(h)h=P_\chi.\]
In the last equality, we used the substitution $h=g^{-1}$.

Next, we have
\be{sumPchi}
\sum_\chi P_\chi = \frac{1}{|G|}\sum_\chi \sum_{g\in G}  \chi(g) g = \frac{1}{|G|} \sum_{g\in G}\big(\sum_\chi \chi(g)\big)g
\ee
The function $\sum_\chi \chi$, as a function of $g$,  is the character of the regular representation (being the sum of all its irreducible characters). The  matrix associated with $g$ 
(as a linear transformation on $\CC[G]$)  is a permutation matrix which transforms the basis $\{ h \, | \, h \in G \}$ to $\{gh \, | \, h \in G \}$. Such a permutation has trace zero for any $g \in G$, except when  $g$ is the identity. In that case $\sum_{\chi}\chi(g)=|G|$, and the right hand side of (\ref{sumPchi}) is equal to the identity. 

{Lastly, we need to show that $P_\chi g P_\chi = \lambda_g P_\chi$, for some $\lambda_g$ depending on $g$, i.e., 
property (\ref{P3}).  }
In fact, we have, since the group $G$ is Abelian,  
$$
P_{\chi} g P_{\chi}=P_{\chi}P_{\chi}g=P_{\chi}g=\frac{1}{|G|}\sum_{h \in G} \chi(h)hg=
\frac{1}{|G|}\sum_{m \in G} \chi(mg^{-1})m=
$$
$$
\frac{1}{|G|}\sum_{m \in G} \chi(m) \chi(g^{-1})m=
\chi(g^{-1})\left( \frac{1}{|G|}\sum_{m \in G} \chi(m) m\right)=\chi(g^{-1})P_\chi, 
$$
as desired. 
\end{proof}

\section{Application to  spin networks subject to symmetries}\label{esempi}

We now apply the above described method to the analysis of the dynamics of 
 two examples concerning networks of spins.  

\subsection{Completely symmetric spin networks}

Consider a network of $n$ {\it identical} spin $\frac{1}{2}$ particles under the control action of a common magnetic field and exhibiting identical ${\it Ising}$ interaction with each other \cite{Xinhua}.\footnote{See also \cite{W} and \cite{GHZ} for interesting quantum states possibly generated by these systems.} We denote by $0\rangle$ and $|1\rangle$ the states of the spin $\frac{1}{2}$ particle, i.e., the two possible eigenstates when measuring the spin in a given direction (e.g., the $z$-direction). Since every spin interacts with every other spin in the same way, we call such networks {\it completely symmetric}.  The state space is $V^{\otimes n}$ where $V = \CC^2$ with the standard inner product 
$\langle \phi|\psi \rangle:= \phi^*\psi$.  Schr\"odinger equation for the dynamics is given by 
 (\ref{SysVec}) with $A=-iH_{zz}$ and $\sum B_j u_j :=-i H_xu_x -iH_yu_y$, where the {\it quantum mechanical Hamiltonians}, $H_{zz}$, $H_x$ and $H_y$,  acting on $V^{\otimes n}$, are given by 
\begin{align}
H_x &= \sum {\bf 1} \otimes \cdots\otimes {\bf 1} \otimes \sigma_x \otimes {\bf 1} \otimes \cdots\otimes {\bf 1}, \label{Ham1} \\
H_y &= \sum {\bf 1}\otimes \cdots\otimes {\bf 1} \otimes \sigma_y \otimes {\bf 1} \otimes \cdots\otimes {\bf 1}, \label{Ham2}\\
H_{zz} &= \sum {\bf 1}\otimes \cdots \otimes {\bf 1} \otimes 
\sigma_z \otimes {\bf 1} \otimes \cdots \otimes {\bf 1}\otimes 
\sigma_z\otimes {\bf 1} \otimes\cdots \otimes {\bf 1}, \label{Ham3}
\end{align}
$u_x$ and $u_y$ represent $x$ and $y$ components of the external (semi-classical) control magnetic field and $\sigma_{x,y,z}$ are the Pauli matrices defined in (\ref{PauliMat}). In (\ref{Ham1}), (\ref{Ham2}) the sum is taken over all the  spins, which are assumed identical, while in 
(\ref{Ham3}) it is taken over all the $n \choose 2 $ pairs   of spins. The group of all permutations on $n$ objects, i.e., the symmetric group $S_n$, acts as a group of symmetries for this system  by permuting the factors in the tensor products:
\begin{equation}
\Pi (v_1\otimes\cdots\otimes v_n) = v_{\Pi(1)}\otimes\cdots\otimes v_{\Pi(n)},  \qquad \forall \, \Pi\in S_n.
\end{equation}
Let $u^{S_n}(2^n):=\big(u(2^n)\big)^{S_n}$. 
The three Hamiltonians (\ref{Ham1}), (\ref{Ham2}), (\ref{Ham3}) commute with the action of 
the symmetric group $S_n$.  
Therefore the dynamical Lie algebra ${\cal L}$ is a subalgebra of $u^{S_n}(2^n)$. The dimension of $u^{S_n}(2^n)$ was calculated in 
\cite{AD} to be ${n+3} \choose n$. In  fact, it  was shown in \cite{AD}, that the dynamical Lie algebra ${\cal L}$ in this case is {\it exactly equal} to $ u^{S_n}(2^n) \cap  su(2^n)$, i.e., $ u^{S_n}(2^n)$ with the restriction that the trace is equal to 0.

Models of this type often represent  crystals of identical equidistant particles. The fact that the particles have the same distance from each other implies that they have the same interaction with each other. 

The GYS's and the associated change of coordinates can be calculated with 
the method of Young tableaux described in subsection \ref{YoungTab7}. Here we calculate the explicit change of coordinates for the case $n=4$. This case is not only the simplest case that was not treated in \cite{AD} but also the highest dimension physically relevant when we consider  spin networks, since symmetry often requires that the spins are equidistant.  Therefore in $3-$dimensional space there are at most  $4$ of them.  In the following we denote by $S_{a_1,a_2,...,a_r}$ the symmetrizer of positions ${a_1,a_2,...,a_r}$ and by $ A_{a_1,a_2,...,a_r}$ the anti-symmetrizer of positions ${a_1,a_2,...,a_r}$, i.e., (cf. (\ref{al})) 
\be{newsymasym}
S_{a_1,a_2,...,a_r}:=\sum_{\sigma \in S_{\{a_1,a_2,...,a_r\}}} \sigma, \qquad  
 A_{a_1,a_2,...,a_r}:=\sum_{\sigma \in S_{\{a_1,a_2,...,a_r\}}} 
\texttt{sgn}(\sigma) \sigma, 
\ee
where $S_{\{a_1,a_2,...,a_r\}}$ is the permutation group of the symbols 
$\{a_1,a_2,...,a_r\}$. We also denote by $V_j$, $j=0,1,2,3,4$, the subspaces of $V^{\otimes 4}$ spanned by states with $j$, $1$'s, so that, for instance, $V_0=\texttt{span} \{ |0000\rangle \}$.

\subsubsection{Young diagram corresponding to the partition $(4)$}  
There is only one Standard Young Tableaux (SYT) corresponding to such a partition given by 
$$
\young(1234) \qquad . 
$$ 
The corresponding   KS-Young Symmetrizer  $P_{_{\tiny{\young(1234)}}}$ coincides with the standard  Young symmetrizer $P_{_{\tiny{\young(1234)}}}^{'} $ (this can be shown by induction to be true for every KS-symmetrizer corresponding to partition $(n)$ for every $n$). The image of $P_{_{\tiny{\young(1234)}}}$ is spanned by the symmetric orthogonal states (for simplicity we omit the normalization factor). 
\begin{equation}\label{phi0}
\varphi_0=|0000\rangle, 
\end{equation}  
\begin{equation}\label{phi1}
\varphi_1=|1000\rangle+|0100\rangle+|0010\rangle+|0001\rangle , 
\end{equation}
\begin{equation}\label{phi2}
\varphi_2=|1100\rangle+|0110\rangle+|0011\rangle+|1001\rangle + |0101\rangle+|1010\rangle, 
\end{equation}
\begin{equation}\label{phi3}
\varphi_3=|0111\rangle+|1011\rangle+|1101\rangle+|1110\rangle , 
\end{equation}
\begin{equation}\label{phi4}
\varphi_4=|1111\rangle. 
\end{equation}

\subsubsection{Young diagram corresponding to the  partition $(3,1)$} 
There are three SYT's corresponding to a partition  $(3,1)$. They are: 
$$
\young(123,4) \, , \qquad \young(124,3) \, , \qquad \young(134,2) \, .  
$$ 
Using the recursive  method of \cite{KS} described in subsection \ref{YoungTab7} we compute the KS-Young symmetrizers and the corresponding bases. 

\begin{itemize}

\item For $P_{_{\tiny\young(123,4)}}$, we get, up to a multiplicative constant, 
$$
P_{_{\tiny \young(123,4)}}=
P_{_{\tiny\young(123)}}
P^{'}_{_{\tiny{\young(123,4)}}}P_{_{\tiny{\young(123)}}}=P^{'}_{_{\tiny{\young(123)}}}P^{'}_{_{\tiny{\young(123,4)}}}P^{'}_{_{\tiny{\young(123)}}}=S_{1,2,3}A_{1,4}S_{1,2,3}, 
$$ 
which applied to $V_0$ and $V_4$ gives zero, while applied to $V_{1,2,3}$ gives the span of $\psi_{1,2,3}$ with 
$$
\psi_1=|1000\rangle+|0100\rangle+|0010\rangle-3|0001 \rangle 
$$
$$
\psi_2=|1100\rangle+|1010 \rangle +|0110 \rangle -|1001 \rangle -|0101 \rangle -|0011\rangle 
$$
$$
\psi_3=|0111\rangle+|1011\rangle+|1101\rangle-3|1110 \rangle.  
$$
Notice that $\psi_3$ is obtained from $\psi_1$ by exchanging the $1$'s with the $0$'s. 

\item For $P_{_{\tiny\young(124,3)}}$, we get, up to a multiplicative constant,
$$
P_{_{\tiny\young(124,3)}}=P_{_{\tiny\young(12,3)}}P^{'}_{_{\tiny{\young(124,3)}}}P_{_{\tiny{\young(12,3)}}}=
P_{_{\tiny\young(12)}}P^{'}_{_{\tiny\young(12,3)}}P_{_{\tiny\young(12)}}P^{'}_{_{\tiny{\young(124,3)}}}P_{_{\tiny\young(12)}}P^{'}_{_{\tiny\young(12,3)}}P_{_{\tiny\young(12)}}=
$$
$$
S_{1,2}A_{1,3} S_{1,2} S_{1,2,4} A_{1,3} S_{1,2}A_{1,3} S_{1,2} =S_{1,2}A_{1,3} S_{1,2,4} A_{1,3} S_{1,2}A_{1,3} S_{1,2}
$$
which applied to $V_0$ and $V_4$ gives zero, while applied to $V_{1,2,3}$ gives the span of $\chi_{1,2,3}$ with 

$$
\chi_1=|1000\rangle + |0100\rangle -2|0010\rangle 
$$
$$
\chi_2=2 |1100 \rangle -2 |0011\rangle + |1001\rangle + |0101 \rangle -|0 11 0 \rangle - |1 0 1 0 \rangle 
$$
$$
\chi_3=|0111\rangle + |1011\rangle -2|1101\rangle.  
$$

\item For $P_{_{\tiny{\young(134,2)}}}$,  we get, up to a multiplicative constant,
$$
P_{_{\tiny{\young(134,2)}}}=P_{_{\tiny{\young(13,2)}}}P^{'}_{_{\tiny{\young(134,2)}}}P_{_{\tiny\young(13,2)}}=
P^{'}_{_{\tiny{\young(1,2)}}}P^{'}_{_{\tiny \young(13,2)}}P^{'}_{_{\tiny\young(1,2)}}P^{'}_{{\tiny \young(134,2)}}P^{'}_{_{\tiny \young(1,2)}}P^{'}_{_{\tiny\young(13,2)}}P^{'}_{_{\tiny \young(1,2)}}=
$$
$$
A_{1,2} S_{1,3} A_{1,2} A_{1,2} S_{1,3,4} A_{1,2} A_{1,2} S_{1,3} A_{1,2} A_{1,2}=
A_{1,2} S_{1,3} A_{1,2} S_{1,3,4} A_{1,2}S_{1,3} A_{1,2},  
$$
which applied to $V_0$ and $V_4$ gives zero, while applied to $V_{1,2,3}$ gives 
the span of $\eta_{1,2,3}$ with 
$$
\eta_1=|1000\rangle - |0100\rangle, 
$$
$$
\eta_2=|1010 \rangle + |1 001\rangle -|0110 \rangle - |0101 \rangle. 
$$
$$
\eta_3=|0111\rangle - |1011\rangle.  
$$
\end{itemize}

\subsubsection{Young diagram corresponding to the partition $(2,2)$} 
There are two SYT's corresponding to a partition  $(2,2)$. They are 
$$
\young(12,34), \qquad \young(13,24).  
$$ 
Using the algorithm in \cite{KS}, we compute the KS-Young symmetrizers and the corresponding bases. 

\begin{itemize}

\item For $P_{_{\tiny{\young(12,34)}}}$, we get, up to a multiplicative constant, 
$$
P_{_{\tiny{\young(12,34)}}}=P_{_{\tiny \young(12,3)}} P^{'}_{_{\tiny\young(12,34)}}P_{_{\tiny \young(12,3)}}=P^{'}_{_{\tiny \young(12)}}
P_{_{\tiny\young(12,3)}}P^{'}_{_{\tiny \young(12)}}P^{'}_{_{\tiny\young(12,34)}}P^{'}_{_{\tiny \young(12)}}
P_{_{\tiny \young(12,3)}}P^{'}_{_{\tiny\young(12)}}=
$$
$$
S_{1,2}A_{1,3}S_{1,2} S_{3,4} A_{1,3} A_{2,4} S_{1,2} A_{1,3} S_{1,2}.
$$
which applied to $V_{0,1,3,4}$ gives zero, while applied to $V_{2}$ gives the span of 
$$
\mu_2=2|1100\rangle +2 |0011\rangle -|0110\rangle -|1010 \rangle -|1001 \rangle - |0101 \rangle 
$$

\item For $P_{_{\tiny \young(13,24)}}$, we get, up to a multiplicative constant, 
$$
P_{_{\tiny \young(13,24)}}=P_{_{\tiny \young(13,2)}} P^{'}_{_{\tiny \young(13,24)}}P_{_{\tiny \young(13,2)}}=
P^{'}_{_{\tiny \young(1,2)}} P^{'}_{_{\tiny \young(13)}} P^{'}_{_{\tiny \young(1,2)}} P^{'}_{_{\tiny \young(13,24)}} 
P^{'}_{_{\tiny \young(1,2)}} P^{'}_{_{\tiny \young(13)}} P^{'}_{_{\tiny \young(1,2)}}=
$$
$$
A_{1,2} S_{1,3} A_{1,2} S_{1,3} S_{2,4} A_{1,2} A_{3,4}A_{1,2}S_{1,3} A_{1,2}= 
A_{1,2} S_{1,3} A_{1,2} S_{1,3} S_{2,4} A_{3,4}A_{1,2}S_{1,3} A_{1,2}, 
$$
which applied to $V_{0,1,3,4}$ gives zero, while applied to $V_{2}$ gives the span of 
$$
\nu_2=|1010\rangle + |0101\rangle -|0110\rangle -|1010 
\rangle. 
$$

\end{itemize}
 
\subsubsection{Structure of the dynamical Lie algebra ${\cal L}$} 
According to the theory developed in this paper, the above change of coordinates transforms the matrices in $u^{S_4}(2^4)$ into a block diagonal form with one copy of $u(5)$ acting on $\texttt{span}\{\varphi_0,\varphi_1,\varphi_2,...,\varphi_4\}$, the so called {\it symmetric states}, three copies of $u(3)$ acting respectively on $\texttt{span}\{\psi_1,\psi_2,\psi_3\}$, $\texttt{span}\{\chi_1,\chi_2,\chi_3\}$, or $\texttt{span}\{\eta_1,\eta_2,\eta_3\}$ and two copies of $u(1)$ acting, respectively,  on $\texttt{span}\{\mu_2\}$ or $\texttt{span}\{\nu_2\}$. Therefore, in the given coordinates, matrices in ${\cal L}=u^{S_n}(2^4) \cap su^{S_n}(2^4)  $ (recall that from the results of \cite{AD} the dynamical Lie algebra ${\cal L}$ is {\it equal} to $u^{S_n}(2^n)$ except for the requirement that the matrices have zero trace) have the form (cf. (\ref{matfor5}))
$$
\begin{pmatrix}
A_{5\times 5} & 0 & 0  & 0 & 0 & 0\cr  
0 & B_{3\times 3} & 0 & 0 & 0 & 0 \cr
0 & 0 & B_{3 \times 3} & 0 & 0  & 0 \cr 
0 & 0 & 0 & B_{3\times 3} & 0 & 0 \cr   
0 & 0 & 0 & 0 & C_{1 \times 1} & 0 \cr 
0 & 0  & 0 & 0 & 0& C_{1 \times 1}  
\end{pmatrix}
$$ 
where $A_{5\times 5}$ is an arbitrary matrix in $u(5)$, $B_{3\times 3}$ is an arbitrary matrix in $u(3)$ and $C_{1\times 1} $ is an arbitrary number in $u(1)$ (i.e., a purely imaginary number), with $Tr(A_{5\times 5})+3Tr(B_{3\times 3})+2C_{1\times 1}=0$. The system is state controllable on each of the invariant  subspaces, that is, it is subspace controllable.   We may calculate the matrices of the restrictions of 
$-iH_{zz}$, $-iH_x$ and $-iH_y$ to the various invariant subspaces and consider control theoretic problems in each subspace.

\subsection{Circularly symmetric spin networks}

Consider  a circular network of identical spin $\frac{1}{2}$ particles interacting via Ising $z$-$z$ interaction but with {\it nearest  neighbor interaction} only. The Hamiltonians modeling 
the interaction with the external magnetic (control) field in the $x$ and $y$ direction are  again given by (\ref{Ham1}) and (\ref{Ham2}).  However   the Hamiltonian modeling the interaction between the particles, $H_{zz}$ in (\ref{Ham3}), has to be replaced by   
\be{Hz4}
H_{zz}^{NN} = \sigma_z \otimes \sigma_z \otimes {\bf 1} \otimes \cdots \otimes {\bf 1} + {\bf 1} \otimes 
\sigma_z \otimes \sigma_z \otimes {\bf 1} \otimes \cdots \otimes {\bf 1}+ \cdots  + {\bf 1}\otimes {\bf 1} \otimes \cdots \otimes {\bf 1} \otimes \sigma_z \otimes \sigma_z +  \sigma_z  \otimes {\bf 1} \otimes \cdots \otimes {\bf 1} \otimes \sigma_z.
\ee
The relevant group of symmetries here is the Abelian subgroup $C_n$  of $S_n$, defined as the group  generated by the circular shift $\{1,2,...,n \} \rightarrow \{n, 1,2,...,n-1 \}$, i.e., the permutation $Z:=(123\cdots n)$, with $Z^n={\bf 1}$.  
 The dynamical Lie algebra ${\cal L}$  is a subalgebra of $u^{C_n}(2^n):=(u(2^n))^{C_n}$. The dimension of $u^{C_n}(2^n)$ is derived in Appendix A and it is given by 
 \be{dimucn}
 \dim u^{C_n}(2^n)=\frac{1}{n}\sum_{m|n} 4^{\frac{n}{m}} \phi(m), 
 \ee
where $\sum_{m|n}$ means we sum over all positive integers $m$ which divide $n$, and $\phi(m)$ is the {\it Euler's totient function} (see, e.g., \cite{ET}) defined as $\phi(1)=1$ and $\phi(m)$ equal to the number of positive integers less than  $m$ which are relatively prime to $m$, if $m > 1$. It is interesting to note that, contrary to what happens  in the example of the previous subsection, the dynamical Lie algebra ${\cal L}$ in this case may be  a {\it proper} Lie subalgebra of $u^{C_n}(2^n)$ (modulo the requirement of zero trace).  Consider, for instance,  the case $n=3$. From formula (\ref{dimucn}) since $\phi(1)=1$ and $\phi(3)=2$, we have 
$$
\dim u^{C_3}(2^3)=\frac{1}{3}\left(4^3\times 1 + 4^1 \times 2\right)=24. 
$$
Therefore  $u^{C_n}(2^n)$ is larger than $u^{S_n}(2^n)$, since the latter    has dimension ${{n+3} \choose n}=20$. On the other hand, for $n=3$, the dynamical Lie algebra generated by $iH_{zz}^{NN}$ in (\ref{Hz4}) and $iH_x$ and $iH_y$ in (\ref{Ham1}), (\ref{Ham2}) is the same as the one generated by (\ref{Ham1}), (\ref{Ham2}) and (\ref{Ham3}) since the Hamiltonian $H_{zz}^{NN}$ in (\ref{Hz4}) coincides with  the 
Hamiltonian $H_{zz}$ in (\ref{Ham3}) in this  case. So the 
dynamical Lie algebra is ${\cal L}=u^{S_n}(2^n) \cap su(2^n)$ in this case because of the result of \cite{AD}. This  has dimension $19$ while $u^{C_n}(2^n) \cap su(2^n)$ has dimension $23$. 

Since $C_n$ is an Abelian group, every finite-dimensional irreducible representation is $1$-dimensional. There are exactly $n$ not equivalent such representations (in the regular representation) which we denote by: $\rho_0, \rho_1,\ldots, \rho_{n-1}$. They are  given by
\begin{gather}
\rho_k:C_n\to GL(1,\mathbb{C})=\mathbb{C}^\times \\
\rho_k(Z^j)=\varepsilon^{kj}
\end{gather}
where $\varepsilon :=\varepsilon_n := e^{2\pi i/n}$ is the $n$-th root 
of the identity.\footnote{For a representation $\rho$, we can write $\rho(Z)$ in Jordan canonical form, in appropriate coordinates. From $\rho(Z)^n=\rho({\bf 1})^n={\bf 1}$ it follows that each Jordan block must be a multiple of the identity, $\lambda {\bf 1}$ with $\lambda$ an $n$-th  root of the identity.} The character associated to the representation $\rho_k$, $k=0,1,2,...,n-1$ is $\chi_k(Z^j):=\texttt{Tr}\rho_k(Z^j)=\varepsilon^{kj}$. Using 
Proposition \ref{GYSAb}, a complete set of  Hermitian GYS's is then given by the following $n$ Fourier sums in the group algebra $\mathbb{C}[C_n]$:
\begin{equation}\label{GYSSC}
P_k = \frac{1}{n} \sum_{j=0}^{n-1} \chi_k(Z^j) Z^j= \frac{1}{n} \sum_{j=0}^{n-1} \varepsilon^{kj} Z^j,\qquad k=0,1,\ldots,n-1.   
\end{equation}

\subsubsection{States and decomposition of the dynamical Lie algebra}

We now want to decompose the Lie algebra $u^{C_n}(2^n)$, which has dimension given in formula (\ref{dimucn}), using the GYS's (\ref{GYSSC}). From this, we deduce the  decomposition of the dynamical Lie algebra ${\cal L}$ for the system of $n$ interacting spin with circular symmetry. 

Let $V=\mathbb{C}^2$ modeling the state of spin $\frac{1}{2}$ systems. States in a basis of $V^{\otimes n}$ are labeled by binary words
 $\underline{a}=a_1a_2\ldots a_n\in\{0,1\}^n$ as follows:
\begin{equation}
\left|\underline{a}\right\rangle = \vec{a}_1\otimes \vec{a}_2\otimes\cdots\otimes\vec{a}_n, 
\end{equation}
where $\vec{0}=\left(\begin{smallmatrix}1\\0\end{smallmatrix}\right)$, $\vec{1}=\left(\begin{smallmatrix}0\\1\end{smallmatrix}\right)$.  According to 
the method of this paper, we need  to describe $\texttt{Im}(P_k)$, for a 
complete set of GYS's $\{P_k\}$. We notice that the space $V^{\otimes n}_T$ defined as the span of states $|\underline{a}\rangle$ with $\underline{a}$  a word of period $T$ (necessarily) dividing $n$,  is invariant under $C_n$ and therefore it is invariant under the action of any element of the group algebra $\CC[C_n]$ including   the GYS's $\{P_k\}$. The period $T$ is the smallest  positive 
integer such that $Z^T(\underline{a})=\underline{a}$.  We have 
\be{imag5}
(P_k V^{\otimes n})=P_k \bigoplus_{T|n}  V^{\otimes n}_T=\bigoplus_{T|n} (P_k V^{\otimes n}_T).
\ee
Consider a general vector $|\underline{a}\rangle$ 
in the  standard basis of $V^{\otimes n}$ and belonging to $V^{\otimes n}_T$. With a
  GYS,  $P_k$, defined in (\ref{GYSSC}), we have 
\begin{equation}\label{eq:cyclic-states-2}
P_k\big(\left|\underline{a}\right\rangle\big)=\frac{1}{n}\sum_{j=0}^{n-1}\varepsilon^{kj}\cdot\left|a_{1+j}a_{2+j}\cdots a_{n+j}\right\rangle
\end{equation}
where the indices of $a_l$ are considered modulo $n$. Since  the 
word $a_1a_2\cdots a_n$ is periodic of period $T$, that is, 
$
a_{1+T}a_{2+T}\cdots a_{n+T} = a_1a_2\cdots a_n. 
$
or $Z^T(\underline{a})=\underline{a}$, in the right hand 
side of \eqref{eq:cyclic-states-2}, we can divide 
the summation variable $j$ by $T$ to get
\begin{equation}
j=Tq+r,\qquad 0\le r<T,\; 0\le q<\frac{n}{T}.
\end{equation}
Thus
\begin{equation}
P_k\big(\left|\underline{a}\right\rangle\big)=\frac{1}{n}\sum_{r=0}^{T-1} \big(\sum_{q=0}^{\frac{n}{T}-1} \varepsilon^{kTq+kr}\big) \cdot \left|a_{1+r}a_{2+r}\cdots a_{n+r}\right\rangle = \frac{1}{n}\sum_{r=0}^{T-1} \varepsilon^{kr}\big(\sum_{q=0}^{\frac{n}{T}-1} \varepsilon^{kTq}\big) \cdot \left|a_{1+r}a_{2+r}\cdots a_{n+r}\right\rangle.
\end{equation}
The quantity in parenthesis can be computed as a geometric series to give 
\begin{equation}
\sum_{q=0}^{\frac{n}{T}-1} \varepsilon^{kTq} = 
\begin{cases}
 \frac{n}{T},& \text{if } \varepsilon^{kT}=1,\\
\frac{(\varepsilon^{kT})^{n/T}-1}{\varepsilon^{kT}-1} = 0,& \text{otherwise,}
\end{cases}
\end{equation}
because  $\varepsilon^n=1$, since  by definition $\varepsilon:=e^{i\frac{2\pi}{n}}$. Using this, we get
\begin{equation}\label{Pkaction}
P_k\big(\left|\underline{a}\right\rangle\big)=
\begin{cases}
\frac{1}{T}\sum_{r=0}^{T-1} 
\varepsilon^{kr}
 \cdot \left|a_{1+r}a_{2+r}\cdots a_{n+r}\right\rangle,
&  \text{if } \varepsilon^{kT}=1,\\
0,&\text{otherwise}.
\end{cases}
\end{equation}
Then $P_k\big(\left|\underline{a}\right\rangle\big)$ is non-zero if and only if $\varepsilon^{kT}=1$, which happens if and only  $n/T$ divides $k$. Therefore $P_k V^{\otimes n}_T$ in (\ref{imag5}) is nonzero only if $n/T$ divides $k$.
\bex{n=4ex} Consider $n=4$, so that $V^{\otimes n}$ is $16$-dimensional. 
In general the possible values of period (dividing $n=4$) are $T=1$, $T=2$, and $T=4$. 
Let us calculate $\texttt{Im}(P_0)$. All $T=1$, $T=2$, and $T=4$ 
are such that $n/T=4/T$, divide $k=0$. We have one state for each 
orbit of $C_4$, which gives $6$ states 
$\frac{1}{4}\sum_{j=0}^3 Z^j |0000\rangle$,  
$\frac{1}{4}\sum_{j=0}^3 Z^j |1111\rangle$, 
$\frac{1}{4}\sum_{j=0}^3 Z^j |1000\rangle$, 
$\frac{1}{4}\sum_{j=0}^3 Z^j |1100\rangle$, 
$\frac{1}{4}\sum_{j=0}^3 Z^j |1010\rangle$, 
and $\frac{1}{4}\sum_{j=0}^3 Z^j |0111\rangle$, 
which span $\texttt{Im} P_0$. For $k=1$ the only possibility is $T=4$, so that $n/T=1$. 
We have the three states: $\frac{1}{4}\sum_{j=0}^3 \epsilon^j Z^j |1000\rangle$, 
$\frac{1}{4}\sum_{j=0}^3 \epsilon^j Z^j |0111\rangle$, $\frac{1}{4}\sum_{j=0}^3 \epsilon^j Z^j |1100\rangle$. For $k=3$ the only possibility is also $T=4$, and we also have three states:  $\frac{1}{4}\sum_{j=0}^3 \epsilon^{3j} Z^{j} |1000\rangle$, 
$\frac{1}{4}\sum_{j=0}^3 \epsilon^{3j} Z^j |0111\rangle$, and $\frac{1}{4}\sum_{j=0}^3 \epsilon^{3j} Z^j |1100\rangle$. For $k=2$ the possibilities are $T=4$ and $T=2$. For $T=4$, we also have three states:$\frac{1}{4}\sum_{j=0}^3 \epsilon^{2j} Z^{j} |1000\rangle$, 
$\frac{1}{4}\sum_{j=0}^3 \epsilon^{2j} Z^j |0111\rangle$, and $\frac{1}{4}\sum_{j=0}^3 \epsilon^{2j} Z^j |1100\rangle$. For $T=2$, we have one state 
$\frac{1}{4}\sum_{j=0}^3 \epsilon^{2j}Z^j|1010\rangle$. Therefore we have 
$\dim(\texttt{Im} P_0)=6$, $\dim(\texttt{Im} P_1)=3$, $\dim(\texttt{Im} P_2)=4$, 
$\dim(\texttt{Im} P_3)=3$, so that $u^{C_n}(2^4)=u(6) \oplus u(3) \oplus u(4) \oplus u(3)$, since all the irreducible representations associated to the GYS, $P_k$, are inequivalent. The dimension of $u^{C_n}(2^4)$ which is equal 
to $6^2+3^2+4^2+3^2=70$ can also be calculated using 
formula (\ref{dimucn}), which 
gives $\frac{1}{4}(4^4+4^2+2\times 4^2)=70$.  
\eex
Generalizing the previous example, we now want to calculate the dimension of $\texttt{Im}(P_j)$, 
which we denote  by $m_j:=\dim (\texttt{Im}(P_j))$, so that,  
\begin{equation}
u^{C_n}(2^n) = u(m_0)\oplus u(m_1)\oplus\cdots\oplus u(m_{n-1}). 
\end{equation}

Consider the set $X_k$ of binary words $\underline{a}$ of length $n$ and with a period $T$ such that $n/T$ divides $k$. Since the cyclic group $C_n$ preserves the period, $X_k$ is invariant under $C_n$. The cyclic group $C_n$ acts on 
$X_k$ by cyclic permutations of the letters. Moreover, as we have seen above, $P_k$ is non zero only on the vector subspace of $V^{\otimes n}$ spanned by the vectors corresponding to the words in $X_k$. Similarly to what done in Proposition \ref{OrbitProp} in the Appendix A, there is a one to one correspondence between the orbits of $C_n$ in $X_k$ and elements in a basis of $\texttt{Im}(P_k)$ given, using (\ref{Pkaction}),   by 
\be{corrorb}
[(a_1a_2\cdots a_n)] \in X_k/C_n \leftrightarrow \frac{1}{T}\sum_{r=0}^{T-1} 
\varepsilon^{kr}
 \cdot \left|a_{1+r}a_{2+r}\cdots a_{n+r}\right\rangle,
\ee
which is independent of the representative chosen for $[(a_1a_2\cdot a_n)]$. In particular $m_k=\dim{\texttt{Im}(P_k)}=|X_k/C_n|$. Using this, we obtain in Appendix A 
\be{formuladime1}
m_k = \frac{1}{n}\sum_{m|\mathrm{gcd}(n,k)} \mathsf{w}(n,k,m) \cdot \phi(m). 
\ee
{Here $\mathsf{w}(n,k,m)$ is the number of binary words $\underline{a}$ of length $n$ which have a period  $T$ such that  $m$ divides $n/T$ and $n/T$ divides $k$. Consider as an example $\mathsf{w}(6,4,2)$. Since $n=6$, possible values for the periods $T$ are $T=1,2,3,6$. For $T=1$, $\frac{n}{T}=6$, which does not divide $k=4$. For $T=2$, $\frac{n}{T}=3$ but $m=2$ does not divide $\frac{n}{T}=3$. For $T=6$, $\frac{n}{T}=1$, but $m=2$ does not divide $\frac{n}{T}=1$. However for $T=3$, we have $\frac{n}{T}=2$. $m=2$ divides  $\frac{n}{T}=2$ and $\frac{n}{T}=2$ divides $k=4$. We count the number of binary words of period $3$ with $6$ elements which are $6$. Therefore $\mathsf{w}(6,4,2)=6$.} In formula (\ref{formuladime1})  again, as in formula (\ref{dimucn}),  $\phi(m)$ denotes the Euler's totient function computed at $m$. 

The following is a case where we are able to calculate the dynamical decomposition of $u^{C_n}(2^n)$ explicitly. 

\subsubsection{The case where $n$ is a prime number}

Suppose $n=p$ where $p$ is a prime number.
If $k=0$ there are two terms in the sum (\ref{formuladime1}), the one corresponding to $m=1$ and the one corresponding to $m=p$. For $m=1$ we can take words of period $T=1$ and $T=p$ which represent all possible $2^{p}$ words. So we have a term $2^{p} \phi(1)=2^p$ in the sum. For $m=p$ we can only take words of period $T=1$, since words of period $T=p$ are such that $n/T=1$ and $m=p$ does not divide $1$. There are only $2$  such words $(000\cdots 0)$ and  $(111\cdots 1)$. Thus we have a term $2 \phi(p)=2(p-1)$ in the sum. Therefore, we have 
$$
m_0=\frac{1}{p}\left(2^{p}+2(p-1) \right)=2+\frac{(2^p-2)}{p}. 
$$
Notice that, for any integer $a$ and prime number $p$, the quantity $a^p-a$ is divisible by $p$, by Fermat's Little Theorem (see, e.g., \cite{Long}).
If $k>0$ then, independently of the value of $k$,  the only possible period  
in the sum (\ref{formuladime1}) is $T=p$ and the only possible value of $m$ is $m=1$. Thus there is only one term in the sum corresponding to all words except the two of period $T=1$. We obtain 
\begin{equation}\label{poi}
m_k=m_0-2=\frac{1}{p}(2^p-2), \qquad 1\le k <  p.
\end{equation}
Consequently,
\begin{equation}\label{genprime}
u^{C_p}(2^n) =
\begin{bmatrix}
u\big(2+(2^p-2)/p)\big) &                      &        & \\
                        & u\big((2^p-2)/p\big) &        & \\
                        &                      & \ddots & \\
                        &                      &        & u\big((2^p-2)/p\big)
\end{bmatrix}
\end{equation}. 
The dimension is equal to
\begin{equation}\label{simpliffor}
\dim u^{C_p}(2^n) = \frac{(2^p+2p-2)^2+(p-1)(2^p-2)^2}{p^2} = 4+(4^p-4)/p
\end{equation}
after simplification. This also agrees with the formula (\ref{dimucn}) for $n=p$,  a prime number.

\subsubsection{The dynamical Lie algebra for a circularly symmetric spin network} 
As we have discussed above, the dynamical Lie algebra 
${\cal L}$ associated with a circularly symmetric network of spin $\frac{1}{2}$ particles may, in general, be a {\it proper} subalgebra of $u^{C_n}(2^n)$. Nevertheless the change of coordinates  we have obtained in this section places ${\cal L}$ in a block diagonal form from which its  structure is easier  to understand. We illustrate this for the case $n=3$. 

Since $n=3$ is a prime number, we can use the simplified 
formula (\ref{simpliffor}) for 
$m_0=\dim(\texttt{Im}(P_0))$, $m_1=\dim(\texttt{Im}(P_1))$, 
$m_2=\dim(\texttt{Im}(P_2))$, and we get $m_0=4$, $m_1=2$, $m_2=2$. From (\ref{Pkaction}) we obtain a formula for an orthogonal basis of 
$\texttt{Im}(P_0)$ which, after normalization, is  given by 
\begin{eqnarray}
\varphi_0:=|000\rangle; \nonumber \\
\varphi_1:=|111\rangle;  \nonumber\\
\varphi_2:=\frac{1}{\sqrt{3}}(|100\rangle+ |010\rangle+ |001\rangle) \nonumber\\
\varphi_3:=\frac{1}{\sqrt{3}} (|011\rangle+ |101\rangle+ |110\rangle).  \label{statiphi}
\end{eqnarray} 
We also obtain a formula for an orthonormal  basis of  $\texttt{Im}(P_1)$ ($\epsilon:=e^{\frac{i2\pi}{3}}$) 
\begin{eqnarray}
\psi_1:=\frac{1}{\sqrt{3}}(|100\rangle+ \epsilon |010\rangle+ \epsilon^2 |001\rangle) \nonumber\\
\psi_2:=\frac{1}{\sqrt{3}} (|011\rangle+ \epsilon |101\rangle+ \epsilon^2 |110\rangle),  \nonumber\\
\end{eqnarray}
and a formula for an orthonormal basis of $\texttt{Im}(P_2)$, 
\begin{eqnarray}
\eta_1:=\frac{1}{\sqrt{3}}(|100\rangle+ \epsilon^2 |010\rangle+ \epsilon |001\rangle) \nonumber\\
\eta_2:=\frac{1}{\sqrt{3}} (|011\rangle+ \epsilon^2 |101\rangle+ \epsilon |110\rangle). \nonumber\\
\end{eqnarray}
By calculating the action of $-iH_{zz}^{NN}$, $-iH_x$ and $-iH_y$ in 
(\ref{Ham1}), (\ref{Ham2}), (\ref{Hz4}) on the above basis, using the fact that $1+\epsilon+\epsilon^2=0$, we obtain, the expression of these operators in the new basis. This  is, $-i\hat H_{zz}^{NN}=\texttt{diag}(-3i,-3i,i,i,i,i,i,i)$, and  
$$
-i\hat H_x:=\left[ 
\begin{array}{c|c|c}
\begin{matrix}
0 & 0 & -i \sqrt{3} & 0 \cr 
0 & 0 & 0 & -i \sqrt{3} \cr
-i\sqrt{3} & 0 & 0 & -2i \cr 
0 & -\sqrt{3}i & -2i & 0 
\end{matrix} & 0 & 0 \\
\hline
0 & \begin{matrix} 0 & i \cr i & 0 \end{matrix} & 0 \\
\hline
0 & 0 &  \begin{matrix} 0 & i \cr i & 0 \end{matrix}
\end{array}
\right], \quad 
-i\hat H_y:=\left[ 
\begin{array}{c|c|c}
\begin{matrix}
0 & 0 &  \sqrt{3} & 0 \cr 
0 & 0 & 0 & - \sqrt{3} \cr
-\sqrt{3} & 0 & 0 & 2 \cr 
0 & \sqrt{3} & -2 & 0 
\end{matrix} & 0 & 0 \\
\hline
0 & \begin{matrix} 0 & -1 \cr 1 & 0 \end{matrix} & 0 \\
\hline
0 & 0 &  \begin{matrix} 0 & -1 \cr 1 & 0 \end{matrix}
\end{array}
\right]. 
$$
The upper left blocks generates any possible $4 \times 4$ skew-Hermitian block, while the $2 \times 2$ blocks are required to be equal, something 
which is not true for general matrices in $u^{C_3}(2^3)$ (cf. equation (\ref{genprime})). Therefore the dimension of the dynamical Lie algebra is $4^2+2^2-1=20-1$, where the $-1$ is due to the fact that the trace has to be equal to zero. In fact such a Lie algebra coincides with the one we would have obtained had we considered the full symmetric group $S_3$ as the symmetry group of the model. From this decomposition we can infer further properties concerning the  {\it subspace controllability} of the system under consideration. We know that the subsystems identified by the vectors $\{\varphi_0,\varphi_1,\varphi_2, \varphi_3\}$, $\{\psi_1, \psi_2\}$, 
$\{\eta_1,\eta_2\}$, are all state controllable. Therefore we have 
controllability for any invariant subspace, i.e., subspace controllability.

\section*{Acknowledgement} Domenico D'Alessandro is supported by the NSF 
under Grant ECCS 1710558. Jonas Hartwig is supported by the Simons Foundation under Grant 637600.  The authors would like to thank Dr. F. Albertini for helpful discussions on the topic of this paper. The authors would also like to thank an anonymous referee for a very careful reading, 
constructive criticism  and several suggestions concerning a first version of this paper.

\section*{Appendix A: Proofs of Formula (\ref{dimucn}) and of Formula (\ref{formuladime1})}

Consider a Lie algebra   ${\cal R}$ which has a basis  ${\cal B}:=\{E_j\}$ which is invariant, as a set,  under  the action of the group $G$, i.e., if $E \in {\cal B}$, $gEg^{-1} \in {\cal B}$, $\forall g \in G$.   
Then we can derive  a basis for ${\cal R}^G$: Let ${\cal O}$ the set of orbits of $G$ in ${\cal B}$ under the above action. 
\bp{OrbitProp} The set of elements 
\be{corre}
\{ \sum_{E_j \in O} E_j \, | \, O \in {\cal O}\}, 
\ee
is a basis of ${\cal R}^G$. In particular, the dimension of ${\cal R}^G$ is equal to the 
number of orbits in ${\cal B}$ under the above described action of $G$ on ${\cal B}$. 
\ep 

\bpr Since the $E_j \in {\cal B}$ form a basis and the orbits are disjoint, then elements $\sum_{E_j \in O} E_j$ in (\ref{corre}) for different orbits $O$ are linearly independent.  Moreover write $F\in {\cal R}^G$ as $F= \sum_{O \in {\cal O}} F_O$ where ${\cal O}$ is the set of orbits  and $F_O$ is a linear combination of elements in ${\cal B}$ 
in the orbit $O$. Since $gFg^{-1}=F$ for each $g \in G$, and each orbit is invariant, we have 
$$
gFg^{-1}= \sum_{O \in {\cal O}} gF_Og^{-1}= F=  \sum_{O \in {\cal O}} F_O,
$$ 
which implies that, for every orbit $O$, and every $g \in G$,  $g F_{O} g^{-1}=F_O$. Write 
$F_O=\sum_j \alpha_j E_j$ where $E_j$ are the elements in the basis ${\cal B}$ which also belong to the orbit $O$, and for some coefficients $\alpha_j$. Fix $j$ and $k$ and a $g \in G$ so that $g$ maps $E_j$ to $E_k$. Such a $g$ always exists because, by definition, the action of $G$ is transitive on its orbits.  By imposing  $g F_{O} g^{-1}=F_O$, using the fact that  the map associated with $g$ is a bijection from the orbit to itself,  we find that $\alpha_j=\alpha_k$. Since, this is valid for arbitrary $j$ and $k$, we find that $F_O$ must be proportional to the elements $ \sum_{E_j \in O} E_j$ in the set (\ref{corre}). 
\epr 
According to the proposition, the dimension of ${\cal R}^G$ can be calculated using the {\it Burnside's orbit counting theorem} (see, e.g., \cite{Rotman}), 
 \be{numorbits}
 \# \texttt{orbits}=\frac{1}{|G|} \sum_{g \in G}|\texttt{Fix}^g|, 
 \ee  
where $\texttt{Fix}^g$ denotes the set of elements fixed by $g$, in ${\cal B}$.

\subsection{Proof of Formula (\ref{dimucn})}
\begin{proof}
By Proposition \ref{OrbitProp} we have
\begin{equation}
\dim u^{C_n}(2^n) = \#\text{orbits}
\end{equation}
where $\#\text{orbits}$ is the number of orbits with respect to the action of $C_n$ on the set of all words of length $n$ in the four symbols $\boldsymbol{1}, \sigma_x, \sigma_y, \sigma_z$.

Recall Burnside counting theorem (\ref{numorbits}) which applied to our case gives: 
\begin{equation}\label{Burnbis}
\#\text{orbits} = \frac{1}{|C_n|} \sum_{g\in C_n} |\text{Fix}^g| \, .
\end{equation}

The cyclic group\footnote{We use the standard convention in group theory denoting by $\langle F_1,F_2,...,F_s \rangle$ the group generated by the set $\{F_1,F_2,...,F_s \}$.} $C_n=\langle Z\rangle$ has a unique subgroup $H_m$ of order $m$ for every positive divisor $m$ of $n$, namely $H_m=\langle Z^{n/m}\rangle$.
Since every element $g$ of $C_n$ generates some subgroup, we can partition $C_n$ into subsets corresponding to which subgroup they generate. Then we get
\begin{equation}\label{Appe56}
\#\text{orbits} =\frac{1}{n}\sum_{m|n}\sum_{\substack{g\in C_n\\ \langle g\rangle = H_m}} |\mathrm{Fix}^g|,   
\end{equation}
where $\sum_{m|n}$ means we sum over all positive integers $m$ which divide $n$. Next we use the fact that a word is fixed by $g$ if and only if it is fixed by the cyclic subgroup $\langle g\rangle$. Thus we get from (\ref{Appe56})
\begin{equation} \label{eq:cyclic-step4}
\#\text{orbits}=\frac{1}{n}\sum_{m|n} \sum_{\substack{g\in C_n\\ \langle g\rangle = H_m}} |\mathrm{Fix}^{H_m}| \, . 
\end{equation}
 Now recall that any cyclic group has many possible generators. In particular if $g$ generates a group $G$ of order $m$, $g^a$ generates $G$ if 
and only if $\mathrm{gcd}(a,m)=1$. Applying this to $G=H_m$, which is 
cyclic of order $m$,  $(Z^{\frac{n}{m}})^a$ generates $H_m$ 
if and only if $\mathrm{gcd}(a,m)=1$. The {\it Euler's totient function} $\phi(m)$ counts  the number of positive integers $a$ less than or equal to $m$ having greatest common divisor $1$ with $m$. Therefore $H_m$ has $\phi(m)$ generators. This means that we can rewrite \eqref{eq:cyclic-step4} as follows:
\begin{equation}\label{toberep}
\#\text{orbits}=\frac{1}{n}\sum_{m|n} |\mathrm{Fix}^{H_m}|\cdot \phi(m) 
\end{equation}

If $m$ is a positive integer that divides $n$ then the number of words of length $n$ in $4$ letters that are fixed by $H_m$ (equivalently, by $Z^{n/m}$) is $4^{n/m}$ because such words are uniquely determined by the first $n/m$ positions, which can be arbitrarily chosen.
This gives us the formula we wanted to show 
$$
\#\text{orbits}=\dim u^{C_n}(2^n) = \frac{1}{n}\sum_{m|n} 4^{n/m} \phi(m).
$$

\end{proof}

\subsection{Proof of Formula (\ref{formuladime1})}
With the same steps as in the previous proof applied to $X_k$ rather then the whole set of words we arrive at (cf., formula (\ref{toberep})) 
\begin{equation}\label{toberep2}
|X_k/C_n|=\frac{1}{n}\sum_{m|n} |\mathrm{Fix}^{H_m}|\cdot \phi(m),  
\end{equation} 
where now the set fixed by $H_m$, $\mathrm{Fix}^{H_m}$ is considered in $X_k$ rather than in the space of all $2^n$ binary words. Recall that $H_m$ is the subgroup generated by $Z^{n/m}$. A word $\underline{a}$ in $X_k$ is fixed by $H_m$ if and only if $Z^{n/m}(\underline{a})=\underline{a}$. This in turn holds if and only $n/m$ is a multiple of the period $T$ of $\underline{a}$. Therefore the words  in $\mathrm{Fix}^{H_m}$ have period $T$ such that $n/T$ divides $k$ and $m$ divides $n/T$. Their number by definition is $\mathsf{w}(n,m,k)$. Moreover in the sum (\ref{toberep2}) $m$ has to divide $n/T$ and therefore $n$, and $n/T$ has to divide $k$, so that $m$ also has to divide $k$. Therefore the nonzero terms are obtained for $m$ at most equal to the greatest common divisor of $n$ and $k$, i.e., $\gcd(n,k)$ which gives formula (\ref{formuladime1}).

\end{document}